\documentclass[alpha-refs]{wiley-article}
\usepackage{color}
\usepackage{ulem}

\usepackage{rotating}[figuresleft]
\usepackage[utf8]{inputenc}
\usepackage{amsmath}
\usepackage{amsfonts}
\usepackage{amssymb}
\usepackage[T1]{fontenc}
\usepackage{graphicx}
\usepackage{natbib}

\papertype{Original Article}
\paperfield{Quarterly Journal of the Royal Meteorological Society }

\title{Domino: A new framework for the automated identification of weather event precursors, demonstrated for European extreme rainfall.}

\author[1]{Joshua Dorrington}
\author[1]{Christian Grams}
\author[2,3]{Federico Grazzini}
\author[4]{Linus Magnusson}
\author[4]{Frederic Vitart}

\affil[1]{Institute of Meteorology and Climate Research (IMK-TRO), Department Troposphere Research, Karlsruhe Institute of Technology (KIT), Karlsruhe, Germany}
\affil[2]{Ludwig-Maximilians-Universität, Meteorologisches Institut, München, Germany}
\affil[3]{ARPAE-SIMC, Regione Emilia-Romagna, Bologna, Italy}
\affil[4]{ECMWF, Shinfield Park, Reading, RG2 9AX, United Kingdom}

\corraddress{Joshua Dorrington PhD, IMK-TRO, KIT, Karlsruhe, 76133, Germany}
\corremail{joshua.dorrington@kit.edu}

\fundinginfo{Karlsruhe Insitute of Technology; German Research Foundation (DFG), Grant/Award Number: FR4363/1-1; Transregional Collaborative Research Centre, Project T2, Helmholtz Young Investigator Group ‘Sub- Seasonal Predictability: Understanding the Role of Diabatic Outflow’ (SPREADOUT; Grant VH-NG-1243)}

\runningauthor{Dorrington et al. 2023}

\begin{document}

\maketitle

\begin{abstract}
%Currently 278 words/300words max
    A growing number of studies have investigated the large-scale drivers and upstream-precursors of extreme weather events, making it clear that the earliest warning signs of extreme events can be remote in both time and space from the impacted region. Integrating and leveraging our understanding of dynamical precursors provides a new perspective on ensemble forecasting for extreme events, focused on building story-lines of possible event evolution. This then acts as a tool for raising awareness of the conditions conducive to high-impact weather, and providing early warning of their possible development. However, operational applications of this developing knowledge-base is limited so far, perhaps partly for want of a clear framework for doing so.
    Here, we present such a framework, supported by open software tools, designed for identifying the large-scale precursors of categorical weather events in an automated fashion, and for reducing them to scalar indices suitable for statistical prediction, forecast interpretation, and model validation.
    We demonstrate this framework by systematically analysing the precursor circulations of daily precipitation extremes across 18 regional- to national-scale European domains. We discuss the precursor rainfall dynamics for three disparate regions, and show our findings are consistent with, and extend, previous findings. We provide an estimation of the predictive utility of these precursors across Europe based on a linear statistical model, and show that large-scale precursors can usefully predict heavy rainfall between two and six days ahead, depending on region and season. We further show how for more continental-scale applications the regionally-specific precursors can be synthesised into a minimal set of indices that drive heavy precipitation. 
    We then provide comments and guidance for generalisation and application of our demonstrated approach to new variables, timescales and regions. 

% Please include a maximum of seven keywords
\end{abstract}

%%%%%%%%%%%%%%%%%%%%%%%%%%%%%%%%%%%%%%%%
%%%%%%%%%%%%%%%%%%%%%%%%%%%%%%%%%%%%%%%%
%%%%%%%%%%%%%%%%%%%%%%%%%%%%%%%%%%%%%%%%
\section{Introduction}\label{sec:intro}
In recent years, the conditioning of surface extreme weather on the large-scale flow, both locally and upstream, has been explored with the hope of obtaining dynamical insights that can identify windows of opportunity and improved guidance for the forecasting of extreme events. High rainfall over North Italy is driven by Rossby wave trains with north American origin \citep{Grazzini2015}, while Atlantic wave-breaking favours Alpine heavy precipitation \citep{Martius2008,Barton2016}, and atmospheric regime variability modulates both alpine serial cyclone clustering \citep{Barton2022,Tuel2022} and European atmospheric river occurrence \citep{Pasquier2019}. Localised temperature and wind extremes are also often associated with hemispheric variability, as revealed by recent studies on compound heatwaves \citep{White2022}, links between cold spells and storm damage \citep{Messori2016}, and the well-documented link between sudden stratospheric warmings and surface cold extremes, modulated by the tropospheric flow \citep{Beerli2019}.

Such large-scale extreme precursors are particularly interesting because they hold the promise of extending forecast predictability. Even in state-of-the-art forecast systems, skill in predicting extreme precipitation, for example, rarely extends even a week ahead \citep{Gascon2023,Leon2023} placing limits on the mitigation strategies available to policy makers and industry stakeholders. However, as forecast skill takes longer to decay at larger spatial scales \citep{Lorenz1969}, large-scale `event-prone' precursors -- which may take the form of particular tropospheric circulation patterns or anomalies in slow-varying earth system modes -- can be expected to be more predictable than the extreme events themselves. As small-scale processes, long-range teleconnections, and orography are imperfectly represented in models, there is no guarantee that a model will always successfully convert the large-scale conditioning into a surface extreme. However, if the conditioning of an extreme event on a large-scale precursor is sufficiently large -- that is, $P(\text{event}|\text{precursor})/P(\text{event})>>1$ -- then such precursors could still be used to offer quantitative improvements in forecast skill by using forecasts of precursors to statistically infer the surface extreme.

Complementarily, from a qualitative perspective a dynamical precursor perspective raises awareness of the necessary conditions for an extreme event rather than focusing on raw model output. This is useful for building forecast story-lines, highlighting important dynamical steps during the unfolding -- or not -- of an extreme even, a perspective which is already popular in the domain of climate projections \citep{Shepherd2018}. As these precursors often occur well before the extreme itself, and may be tied to predictability barriers \citep{Gonzalez-Aleman2022,Oertel2023}, such an approach can help explain `jumpiness' in situations of high forecast uncertainty, increasing end-user trust \citep{Richardson2020}.

Despite these advantages, there has been relatively little operational use of large-scale flow precursors thus far to improve operational forecasts. One notable exception is the use of regime approaches \citep{Ferranti2015}, which have been shown to boost precipitation forecast skill \citep{Mastrantonas2021}. Here, the approach is typically top-down; regimes are identified in large-scale fields such as geopotential height, and the impacts on surface weather in each regime are diagnosed \citep{Grams2017,VanDerWiel2019}. Other large-scale modes such as the MJO, ENSO or stratospheric polar vortex are also integrated into statistical modelling, especially at seasonal lead times but are less commonly employed in the extended range. Bottom-up regime approaches exist, as in \citet{Bloomfield2021} which look for the dominant flow regimes that explain the most variability in variables of interest (in ibid, energy generation). Of course, there is no guarantee that any pre-established dominant variability modes cleanly project upon a generic extreme event. In fact, if such an event is comparatively rare, as an extreme is by definition, then any predictability coming from dominant variability modes must necessarily be highly diffuse and associated with only minor modifications in occurrence probability. \citet{AghaKouchak2022}, in the context of drought prediction, advocated for the bottom-up, event-centric approach of identifying the precursors of specific event types, rather than `thinking platonically', i.e. thinking in terms of idealised modes of variability. 

In this paper we lay the groundwork needed to make bottom-up event-centric precursor analysis a routine technique in the toolbox of the operational meteorologist, the applied scientist or the forecast end-user. In section 2 we introduce our analysis framework for identifying useful precursors, `Domino'\footnote{For the curious, the name `Domino' was inspired by the imagery of a row of dominoes falling in causal sequence, which serves as a (admittedly optimistically deterministic) metaphor for the sequence of meteorological developments that increase the likelihood of a weather event of interest.}, supported by a corresponding Python implementation, a fully-documented api and clear worked examples of different use-cases. 

In the rest of the paper we apply this new framework to analyse the precursors of daily rainfall extremes across Europe. Extreme weather events in Europe are responsible for billions of euros in economic damage and the loss of hundreds of lives every year \citep{Hunt2011,EEA2022}, with 40\% of these deaths between 1950-2006 a result of flash floods. Such short-timescale and spatially-localised rainfall extremes are particularly challenging to forecast well, especially in regions with complex topography.  Our analysis serves as a demonstration of the precursor approach, but also, by providing the first systematic analysis of national-scale precipitation drivers, as an end in its own right. 

In section 3 we discuss the data and event definitions used, the eighteen spatial regions we choose, and our selection of Domino parameters. In section 4 we will discuss the precursor dynamics of three archetypal regions: North Italy, West Iberia, and North Europe which cover Mediterranean, Atlantic and continental climates respectively, and demonstrate consistency with prior knowledge as well as note a number of more novel observations. In section 5, we will analyse the large-scale precursor patterns in wind, geopotential height, and vapour transport derived for all 18 regions, and discuss commonalities and variations. We will estimate the predictive skill of these precursors using logistic regression, and investigate regional and seasonal differences. In section 6 we show how the regionally-specific precursor indices we have computed can be synthesised for larger scale considerations into a minimal set of European rainfall modes' using partial least squares regression, and we use these to characterise regional variations in rainfall dynamics.
Finally in section 7, we discuss our conclusions and provide suggestions for the generalisation and extension of this work, and how the framework can be best used for directly applied work.

\section{Domino}\label{sec:domino}

\begin{figure}
    \centering
    \includegraphics[width=0.9\linewidth]{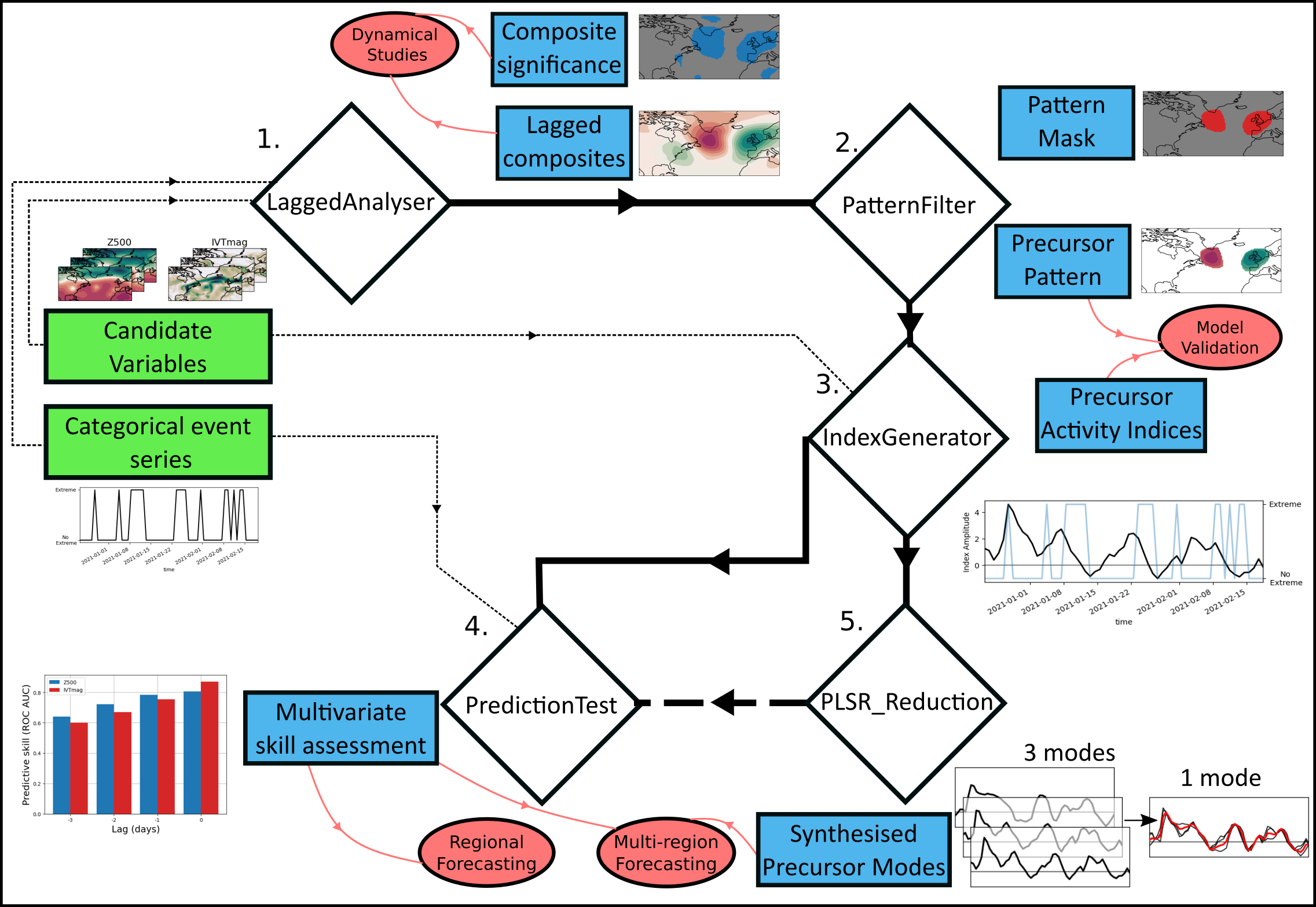}
    \caption{A schematic diagram of the Domino workflow. Inputs are shown in green boxes, key analysis steps in white diamonds, output datasets in blue boxes, and different applications that could make use of output data are indicated by red ovals. See accompanying main text for full description. Graphics show visual examples of each data type.}
    \label{fig:domino_schematic}
\end{figure}

The aim of the Domino framework is to automatically identify the predictive precursors of categorical, meteorological events. 
Comprehensive documentation and discussion of the full range of workflow parameters and customisation options can be found in the Domino github repository \verb|https://github.com/joshdorrington/domino|

In section \ref{sec:methods} we define concretely the parameters and analysis used in this paper for our proof-of-concept application to European extreme rainfall, while in this section we summarise the general purpose and motivation behind each analysis step.

The framework, sketched schematically in figure \ref{fig:domino_schematic}, is based on two types of input: `events', defined as a categorical time-series indicating the class of event at each time-step, and `variables': time-series of candidate precursors which may play a role in influencing the event and which may be time series of scalars, vectors, fields, or higher-dimensional variables. Our overarching goal is to provide a simple workflow that can replicate, semi-automatically, previous analyses of extreme event precursors, and convert the results into a form that can be used for event prediction. In this way we aim to operationalise current academic insights, and streamline the process of obtaining new insights.

\begin{description}
  \item[1. Compute lagged composites] Implemented via the LaggedAnalyser class, time-lagged composites of the input variables are calculated with respect to each event category. Lag-dependent background-states and/or seasonal cycles can be computed and subtracted. Synthetic event time-series can be provided or generated automatically to serve as the basis for bootstrap significance testing. As such compositing approaches are a valuable analytical tool used in many studies of extreme event precursors, long-range teleconnections and atmospheric dynamics, we anticipate the clean, efficient implementation provided here will be of interest to many geoscientists.
  
  \item[2. Refine precursor patterns] Implemented via the PatternFilter class, a number of filters can be applied to precursor composites of multi-dimensional variables, iteratively building a Boolean mask which can be used to focus on features of interest. The motivation is to replicate in an automated way the subjective assessment of interesting features a scientist might make when looking at a composite field. Provided functions include filters for large-amplitude, spatially extended and statistically significant patterns. 
  
  \item[3. Generate precursor indices] Implemented via the IndexGenerator class, each precursor pattern can be reduced to scalar time-series of `precursor activity', by projecting anomaly fields onto the masked pattern. Such indices are the automatically-generated, event-centric equivalents of, for example, the well-known NAO, QBO or nino3.4 indices defined as spatial averages or differences of spatial averages over particular regions, and which provide a valuable scalar summary of important Earth-system variability. As well as computing precursor activity indices from scratch, updating an index using new variable time series and precomputed composites is supported, making it easy to calculate online updates of index time series, or project indices into a different dataset, such as in model data. Although not explored in this work, there is potential for using these indices to validate the event-centric dynamics of models much in the same vein as model assessments of ENSO \citep{Fredriksen2020} and the NAO \citep{Simpson2020}.
  
  \item[4. Evaluate index predictivity] Implemented via the PredictionTest class, this step provides an estimate of the predictive power of scalar indices for a categorical event. Uni- or multi-variate combinations of predictor indices can be specified, as can the statistical model to use, the cross-validation strategy and the skill score of interest. This can be used for feature selection, allowing for a subset of most predictive indices to be taken forward for additional analysis, or as in this paper, to compare the predictability of precursors in different regions and seasons. 
  
  \item[5. Reduce index dimensionality] Implemented via the PLSR\_Regression class, a set of scalar indices can be reduced to a lower-dimensional set via partial-least-squares regression, a statistical model which  maximises the covariance between the reduced modes and a set of target variables. In this way, we can reduce the dimensionality of our scalar indices in an event-focused way which is well suited to prediction problems. This auxilliary functionality is useful in specific cases where precursors are desired which balance predictive power over different event types or across regions, as discussed in section \ref{sec:plsr}. The output PLSR modes can then also be passed to PredictionTest, just as can any other scalar index. 
\end{description}

The software implementation is built on top of the popular Python package xarray \citep{Hoyer2017}, which provides support for meta-data rich, multi-dimensional array manipulation. Domino makes heavy use of its DataArray and Dataset classes to store and process data, which support conversion both to and from Pandas Dataframes and iris Cubes, providing cross-compatibility with a range of data pipelines. Implementations of logistic regression and PLSR are taken from scikit-learn \citep{Pedregosa2011}. Via xarray, much of the workflow supports parallel and lazy computation via Dask \citep{Rocklin2015}, allowing large datasets to be processed quickly.

In this paper we focus on the case where the events have two classes, e.g. case the occurrence or not of extreme rainfall, but all the algorithmic steps of Domino equally apply to multi-class applications; terciles, deciles, events of multiple type, etc. Here we only apply Domino to precursors which are 2D continuous lat-lon fields, but arbitrary dimensioned continuous precursors can be handled, as can categorical scalar precursors. In the latter case, occurrence frequency for each category is automatically computed instead of a mean composite.

\section{Data and Methods}\label{sec:methods}

\subsection{Data}\label{sec:methods_data}
Both events and precursors are identified in ERA5 reanalysis data \citep{Hersbach2020} over the 43-year period 1979-2021. Precipitation data was downloaded as 3-hourly accumulations at 0.25 deg. resolution, and then accumulated to daily data. Four candidate precursor fields were chosen to capture the large-scale circulation: 500 hPa geopotential height (Z500), 850hPa zonal wind (U850), 300hPa meridional wind (V300), and the magnitude of the column integrated vapour transport (IVTmag). These were downloaded as 12:00 instantaneous values at 1 deg. resolution.

\subsection{Region definition}\label{sec:methods_region}
\begin{figure}
    \centering
    \includegraphics[width=0.6\linewidth]{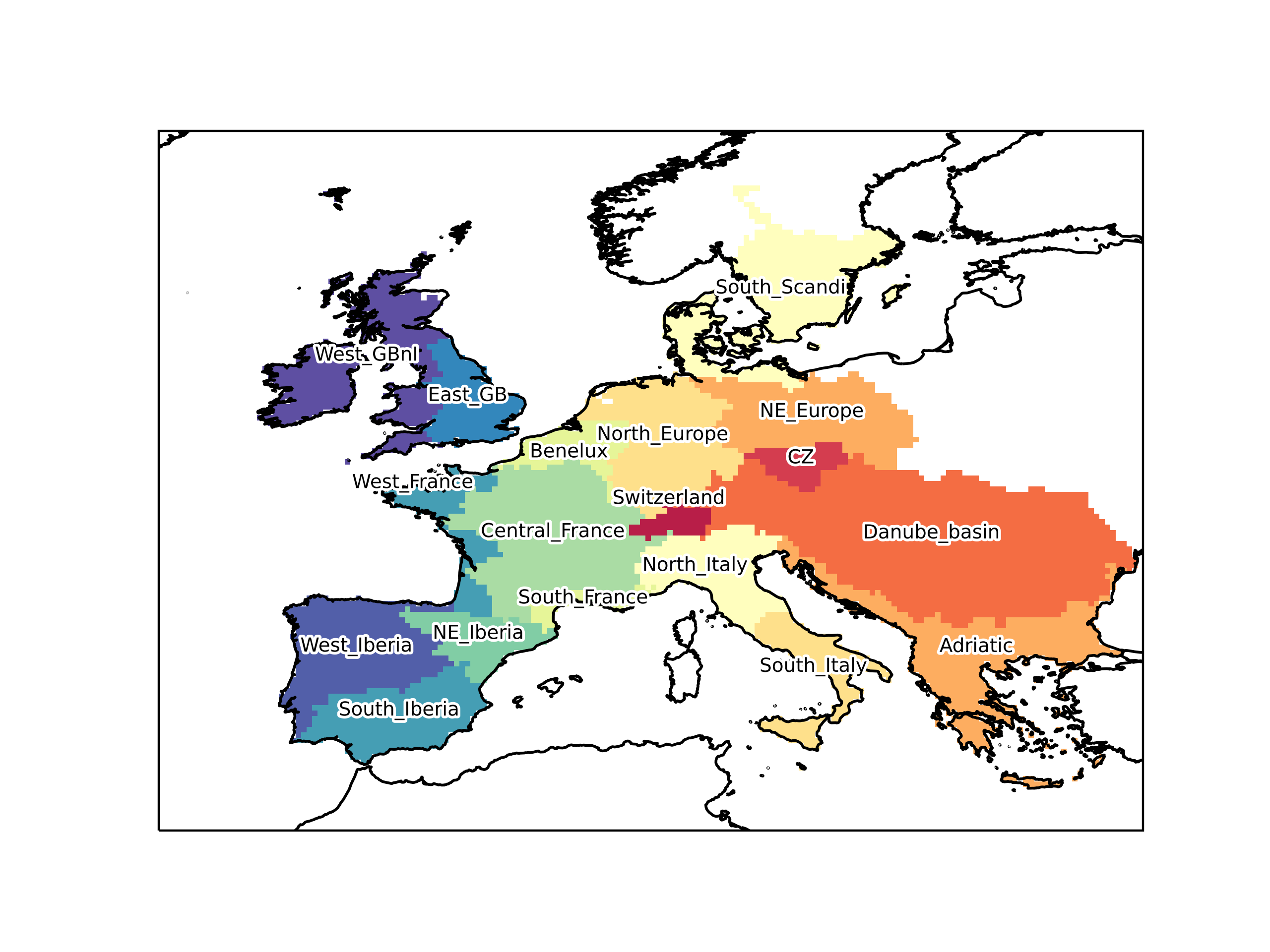}
    \caption{18 regions used in this paper to define spatially-aggregated precipitation extremes. Regions are based on catchment data taken from \citet{Lehner2013}, with modifications to account for orographic impacts on the prevailing circulations.}
    \label{fig:regions}
\end{figure}

Spatially-averaged daily rainfall was defined as the cosine-latitude weighted averages of precipitation over each of the eighteen European regions shown in figure \ref{fig:regions}. The boundaries for these regions were based on the HydroSHEDS drainage-basin product \citep{Lehner2013}, which avoids the mixing of hydrologically distinct areas that would result from a simple box-averaging approach. Basins were in some cases merged or split in a subjective fashion based on consideration of meteorological factors such as the position of mountain ranges and prevailing winds, or to merge very small regions. 

The areas, chosen to demonstrate the Domino workflow, are not of equal area or population and so are therefore not optimised for direct application. Rather, they serve to collectively cover the majority of Europe in a way that respects the underlying hydrometeorology, and the analysis of their precursors can then inform further studies of particular refined regions.

For each season and region, a categorical heavy rainfall event time series was computed. Events were defined as daily accumulated rainfall exceeding the region's seasonal 90th percentile. This relative definition ensures an equal number of samples in all samples and regions, with the drawback that in some cases such events will not always be high-impact. Supplementary figure 1 shows the exact threshold for each case and region.

\subsection{Precursor computation}\label{sec:methods_precursor}

\begin{figure}
    \centering
    \includegraphics[width=\linewidth]{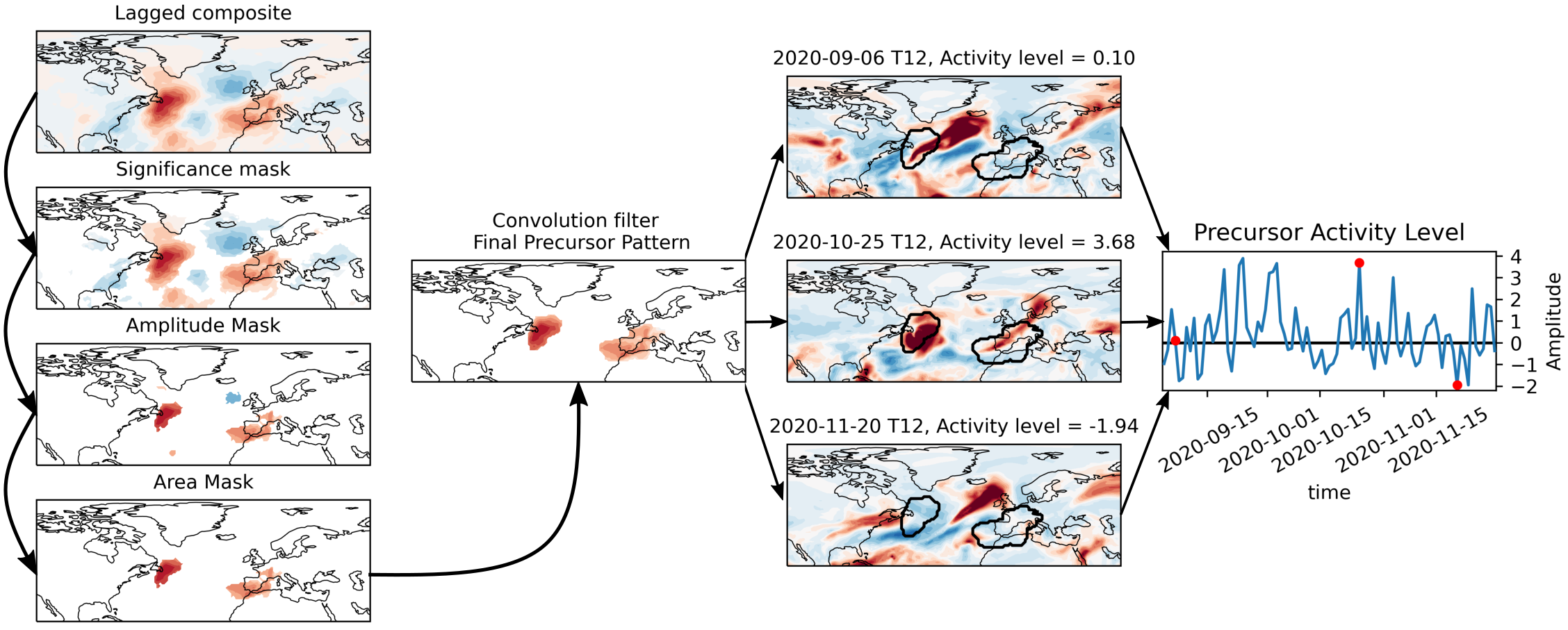}
    \caption{A schematic indicating how lagged composites are transformed into precursor patterns and then into precursor indices, demonstrated for SON IVTmag precursors 2 days prior to North Italian extreme rainfall events. The initial lagged composite is masked based on statistical significance, anomaly amplitude, and spatial extent of unmasked regions, then padded with a 5-point 2D-convolution. A dot-product between this pattern and the daily IVTmag anomaly field discriminates between days that closely resemble the precursor flow and those that do not. When applied to all days in the IVTmag dataset, a daily precursor activity index is produced. The three daily fields used as examples are marked with red dots in the resulting precursor activity time series.}
    \label{fig:precursor_computation}
\end{figure}

The specifics of how precursor patterns and activity indices were computed are detailed here, and are accompanied by schematic figure \ref{fig:precursor_computation}. Each of the precursor variables (Z500, IVTmag, U850, and V300) were restricted to the extended north Atlantic region [20N-80N,140W-100E], and deseasonalised at a gridpoint level, by calendar date: the mean value of each variable on the same day of the year over the period 1979-2021 was smoothed with a 30-day rolling mean, and subtracted from the daily fields. 

For each season and region, lagged mean composites of the precursor variables were computed at daily lags from 7 to 0 days before the corresponding heavy rainfall event series. As heavy rainfall events were defined as exceedance of the 90th percentile threshold in each case, this represents composites of 385-395 individual days.

In order to assess the statistical significance of precursor anomalies, a bootstrap approach was used. To account for any autocorrelation in heavy rainfall days, 400 synthetic event indices were generated by fitting a Markov chain to the observed event series and resampling from it. Composites of each variable were computed with respect to each synthetic event series, and significance was defined at a gridpoint basis as a value below/above the 10th lowest/highest bootstrap composite (2-sided p=0.05). A Holm-Bonferroni correction was then applied to each variable and lag separately \citep{Holm1979}, to reduce the false-positive rate. Composites at all time lags were compared to a single set of bootstrap composites computed at lag 0 for reasons of computational efficiency. This amounts to assuming an approximately stationary climate over a 10-day period, which is reasonable.

Lagged composites are transformed into `precursor patterns' by applying a number of filters to the Boolean significance field, and then using it to mask the composite. This procedure results in statistically robust spatial anomaly patterns with large amplitudes, smooth edges and without isolated unmasked points. Concretely, in addition to masking gridpoints with statistically insignificant anomalies, gridpoints with anomalies <0.25x the gridpoint climatological standard deviation were masked. Groups of unmasked gridpoints that comprise a connected area of less than ~250,000 km$^2$ (i.e. equivalent to 20 1x1 deg equatorial gridpoints) were also masked. Finally, all unmasked gridpoints were convolved with a 5-point lat-lon square: all gridpoints within 2 degrees of an unmasked gridpoint were unmasked. 

These precursor patterns are then used to compute precursor activity indices. A dot-product is computed between each deseasonalised variable time series and each corresponding precursor pattern, and then spatially summed, using cosine-latitude weighting. This produces a scalar index which is standardised to 0 mean and standard deviation 1. 

\subsection{Skill assessment}\label{sec:methods_skill}

\begin{figure}
    \centering
    \includegraphics[width=\linewidth]{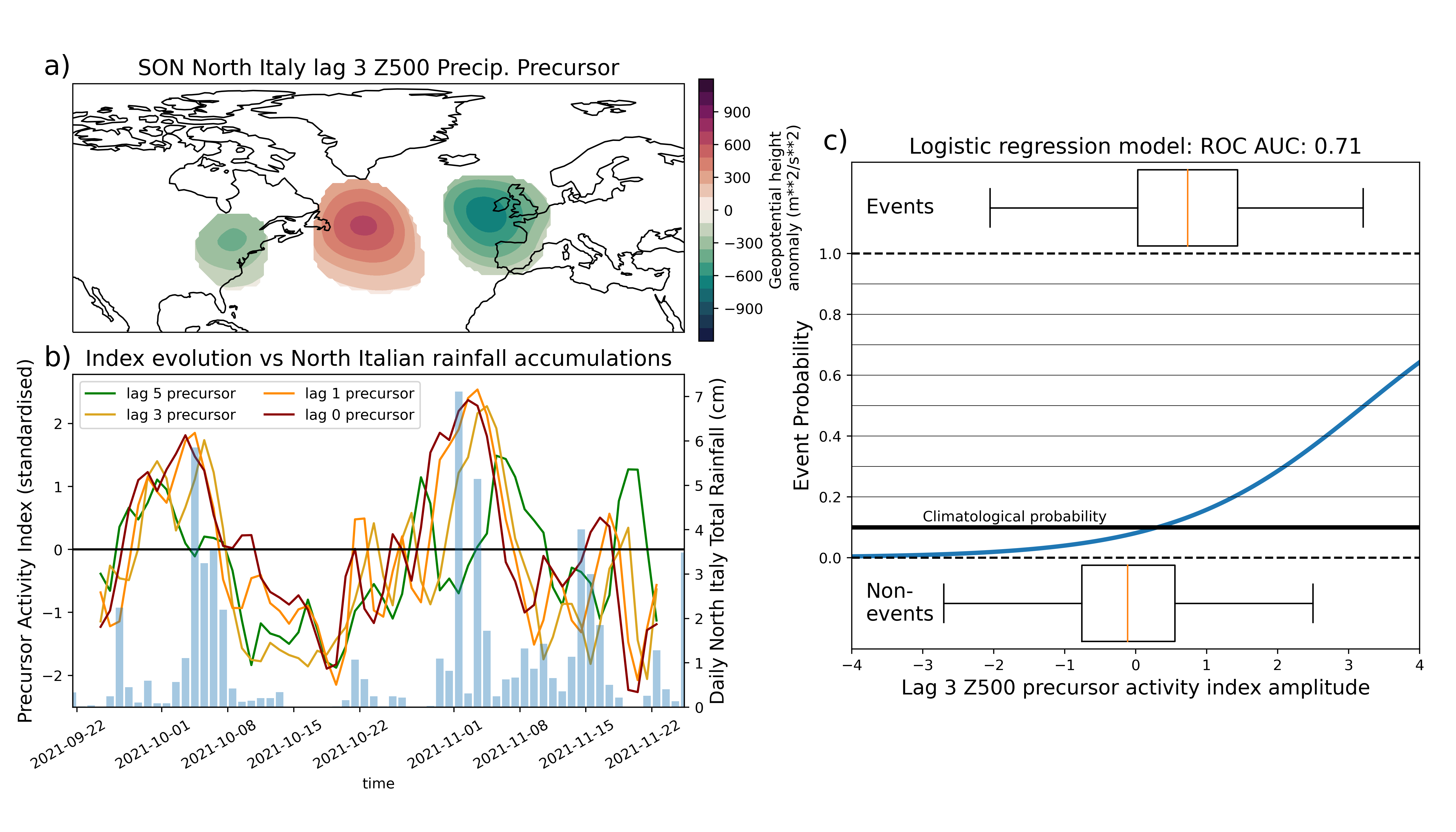}
    \caption{A schematic illustrating how the predictive power of precursor activity indices is estimated in a reanalysis framework. For each time lag, region, season and variable a precursor pattern is defined (a). Each pattern corresponds to a time-varying activity index which captures to what degree the current flow resembles the pattern. These indices co-vary with rainfall to a greater or lesser extent (b)). We fit a logistic regression model with precursor indices for all lags >=$l$ for a given season, region and variable, in order to predict categorical heavy rainfall occurrence, and quantify skill using the ROC AUC score.}
    \label{fig:skill_schematic}
\end{figure}

To characterise the potential predictability of heavy rainfall precursor activity indices, we use a logistic regression approach. Stratifying by season, region and precursor variable, in each case our target variable is the corresponding ERA5 Boolean heavy rainfall event index, where we aim to predict the probability of an event occurring. The schematic figure \ref{fig:skill_schematic} illustrates how this is done.
For each time lag, season, region and large-scale variable we have one precursor pattern (fig \ref{fig:skill_schematic}a). Each index is offset in time according to its lag, i.e. a 4-day precursor index calculated from data for 1st Jan 2000 will be used to predict for 5th Jan 2000. We train a model for each lag $l$ between 7 and 0 by using all precursor indices that have a lead time $\leq l$: thus the 7-day model is a univariate model using only the lag 7 precursor index, while a 1-day forecast uses all precursors except for the lag 0 precursor. 
Each logistic regression model was cross-validated: each year of data was sequentially excluded, and the model was fitted on the remaining 42 years, then used to predict the excluded year. The probabilistic ROC Area Under the Curve (ROC AUC) skill score \citep{Hanley1982} was computed for each cross-validation, and the mean value is reported. This provides a much more realistic estimate of the potential predictability achievable by a perfect forecast than simply fitting a model to the entire event time series.

\subsection{Partial least squares regression}\label{sec:methods_plsr}

Dimensionality reduction is a common task in atmospheric data analysis, with principal component analysis (PCA) being the most popularly used approach. PCA aims to find the N-dimensional linear subspace of an initially M-dimensional dataset which captures the greatest degree of variance. Partial least squares regression (PLSR) is similar, in that it aims to reduce dimensionality while maximising explained variance, but now finding the N-dimensional linear subspace of an M-dimensional dataset $X$ which maximises the cross-covariance with a second L-dimensional dataset $Y$. That is, it estimates the low-dimensional projection of $X$ which is best at predicting the various target variables of $Y$. For the mathematical formalism see \citet{Abdi2010} and for algorithmic implementation see \citet{sklearn-plsr}. It is closely related to canonical correlation analysis, which has been previously applied to meteorology in, e.g., \citet{Maldonado2013}. 
For our example, we demonstrate how PLSR can be applied to SON Z500 precursor patterns for different regions and time lags, to obtain a minimal set of modes that balance week-ahead heavy rainfall predictability across Europe. This is described in more detail in section \ref{sec:plsr}.

\section{Dynamics of rainfall precursors}\label{sec:dynamics}
\begin{figure}
    \centering
    \includegraphics[width=\linewidth]{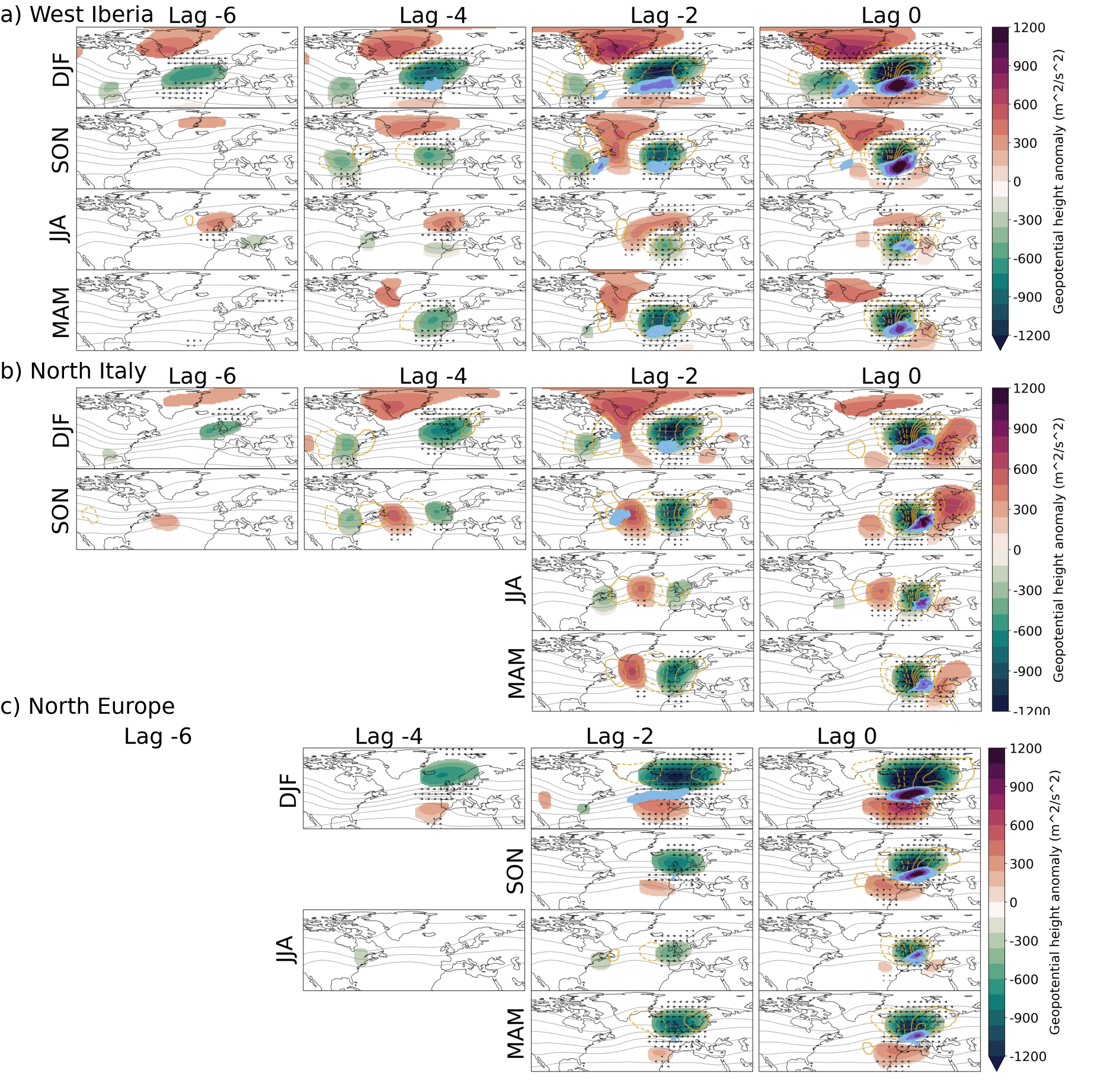}
    \caption{Precursor patterns 6, 4, 2 and 0 days before heavy rainfall events in West Iberia (a,), North Italy (b) and North Europe (c), as a function of season and computed from ERA5 reanalysis data. Red-to-green shading indicates Z500 anomalies as shown by the colorbar. Grey contours show full field Z500 for levels with 1e4 spacing between 5.2e5 and 5.7e5 $m^2/s^2$. Gold contours show V300 anomalies at 4$m/s$ intervals ranging from -16 to 16 $m/s$, and excluding the 0 contour. Blue-purple shading shows positive IVT magnitude anomalies at the 50, 80, 110,and 140 kg$m/s$ levels. Arrows indicate anomalies of the U850 wind, and should be interpreted qualitatively. 
}
    \label{fig:synth_plot}
\end{figure}

Having discussed our approach and methodology in general terms, we now apply precursor analysis to European daily heavy rainfall extremes. We begin, in this section, by analysing the multivariate precursor patterns for North Italy, West Iberia and North Europe. Figure \ref{fig:synth_plot} shows precursor patterns for the three regions, 6, 4, 2 and 0 days before heavy rainfall events and across seasons. For each of our 4 large-scale variables we have a unique precursor pattern for each combination of time lag, season and region. Z500 precursors indicate the anomalous trough and ridge structure associated with events, V300 indicate the upper-level wave dynamics, U850 the near surface jet, and IVTmag the role of anomalous moisture transport. Collectively they therefore capture most of the important mid-latitude driving dynamics and allow a holistic understanding of each event type.

%\begin{sidewaysfigure}
%    \centering
%    \includegraphics[width=\linewidth]{final_figs/fig6.png}
%    \caption{Precursor patterns prior to heavy rainfall events in West Iberia (a), North Italy (b) and North Germany (c), as in figure \ref{fig:synth_plot_small} but now shown more comprehensively, for all lags between 7 and 0, and for all seasons.}
%    \label{fig:synth_plot}
%\end{sidewaysfigure}

\subsection{West Iberia}\label{sec:dynamics_WI}
The `West Iberia' region consists of Portugal and coastal Northern Spain, as well as the inland basins of the Douro and Tagus rivers. Precursors extend further back in time in this region than any other considered, as we will discuss in section \ref{sec:pred}.
In this mountainous domain, the majority of rainfall occurs in Autumn and Winter \citep{Mora2020}, and while Summer rainfall is less frequent and generally dominated by meso-scale processes, the majority of JJA coastal extreme rainfall events are driven by atmospheric rivers \citep{Ramos2018}. An important flood driver in MAM which we do not consider here, is snow-melt. In general, heavy daily precipitation in this region is known to be dominated by the latitudinal shifting of the jet stream which, when in its southerly state, can guide cyclones and atmospheric rivers to this area \citep{Ramos2015}, and is associated with the negative phase of the NAO \citep{Eiras-Barca2016,Salgueiro2013} and cut-off lows to the West \citep{Eiras-Barca2018}.

This pre-existing knowledge is reaffirmed by the large-scale West Iberian precursor patterns shown in Figure \ref{fig:synth_plot}a). Robust precursors extend back 6 days in all seasons, demonstrating a substantial conditioning of heavy rain events here on the upstream Atlantic state. In DJF, precursor in the Z500 and U850 fields indicate a southerly jet deflection, accompanied by a negative-NAO-like geopotential height anomaly, is typical 6 days before a rainfall event. Both wind and geopotential anomalies strengthen through to lag 0. This extended persistence of the precursor flow is consistent with the persistent, trimodal regime dynamics of the wintertime Atlantic jet \citep{Woollings2010,Parker2019}. From lag 4 onwards a clear zonally oriented IVT precursor emerges to the west of Iberia, indicating increased atmospheric river frequency, and the need for not only a displaced jet but also anomalous moisture transport to induce an extreme. While the U850 anomaly is initially confined to the east Atlantic, a secondary trough forms over the south-east US from lag 6, which induces its own local U850 anomaly and a diagonally oriented IVT precursor that suggests a secondary source of moisture flowing from the west to the east Atlantic. While the early precursors are predominately zonal, large V300 anomalies emerging from lag 4 onward indicate the formation of an upper-level trough over the domain and a possible role for wave breaking dynamics for triggering local trough formation as discussed in \citep{Santos2018}. 
In SON, dynamics are similar, but now less zonally symmetric with clearer upper-level wave features, and a more confined trough and associated deep ridge over the mid-Atlantic. Here the role of the secondary cyclone is comparatively more pronounced, perhaps a requirement given the overall weaker moisture transport associated with a weaker low-level jet. In MAM and JJA there are less pronounced large-scale precursors, but a substantial amount of non-local conditioning can still be seen including low pressure over the US, high pressure over Greenland and East Atlantic jet anomalies as well as the high IVT anomalies indicative of increased atmospheric river frequency. The lag 0 Z500 precursor indicates cut-off lows are dominant in Summer, while troughs are more common in Autumn, in agreement with \citet{Pino2016}, which analysed the synoptics of the 24 most severe Iberian floods in the past 150 years. 
The upstream trough over the Panhandle, visible in U850 and Z500 precursors for SON and DJF 2-4 days prior to heavy precipitation, is also associated with a known dynamical pathway uncovered only very recently: \citet{Leeding2023} have showed DJF cold air outbreaks over this same US region are associated with increased Iberian precipitation 0-5 days later, and that this trough pattern is associated with further southward jet deflection. Our results suggest there may also be a role for this teleconnection in SON, something not explored in ibid.

\subsection{North Italy}\label{sec:dynamics_NI}
The North Italian domain suffers from extreme flash flooding events, and is well known to possess substantial conditioning on the large-scale flow and Atlantic Rossby wave activity \citep{Grazzini2015,Grazzini2020}. Highly mountainous, and shielded from direct westerly flows by the French alps, southerly winds are needed to bring large moisture transport into the region, necessitating some level of upstream wave activity.
The precursor patterns (Figure \ref{fig:synth_plot}b)) reveal a sharp distinction between the non-local conditioning of SON and DJF events, which stretch back to lag 6, and MAM and JJA events, which only emerge at lag 2. Despite this, precursor patterns for the non-winter seasons are qualitatively rather similar, all showing zonally oriented wave trains with large-scale moisture transport over Southern Europe. For SON, a pre-existing weak ridge over the west Atlantic interferes constructively with the upstream Rossby wave, which becomes clear by lag 4, and is centred at $\sim$45N. From here the wave amplifies and shows slow eastward movement. High IVT at lag 2 in the west Atlantic may indicate the increased occurrence of extratropical cyclones recurving north-east, which can amplify wave growth and induce Mediterranean cyclone development \citep{Grams2011,Raveh-Rubin2017}. The trough to the west of Italy is fully developed by lag 2, with associated southern European IVT anomaly. Even in JJA, when convective rainfall would be expected to dominant, high IVT over North Italy is seen, likely as a result of unstable moist air from the Mediterranean advecting and rising over the steep orography,  as described by \citep{Grazzini2020a}.
In DJF, the picture is quite different, with precursors showing a Greenland blocking/NAO- precursor pattern emerging, interacting with an upstream wave train, and leading to a more tripolar high latitude circulation anomaly up until lag 2. High geopotential heights around the European trough indicate a cutoff structure, and there is a sharp decline in the amplitude of the Greenland blocking anomaly on the day of the North Italian extreme rainfall. This break down of the blocking and associated regime transition implies elevated levels of wave breaking over western Europe \citep{Masato2012}, which may aid in deepening and localising the trough, and so amplifying southerly moisture transport. 
For the North Italian case, the Z500, V300 and IVTmag precursor indices have been shown in \citet{Grazzini2023} to increase the predictive power of a hybrid dynamical-statistical prediction model, indicating a capability to not only reproduce the known dynamics, but to transform those into a usable source of skill.

\subsection{North Europe}\label{sec:dynamics_NC_Europe}
The `North Europe' domain consists of the catchments of the Ems and Weser rivers, as well as the northern portion of the Rhine and Meuse catchments, including large parts of Germany, eastern Belgium and the northern Netherlands. In contrast to the other two domains considered in this section, this domain which excludes the German alps and the Ore mountains to the south and east respectively, features relatively flat terrain and exhibits a more continental climate. Although the heavy rainfall threshold is comparable in all seasons (~6mm/day), there are notable seasonal variations in the precursor patterns (figure \ref{fig:synth_plot})c). Without strong orographic constraints or direct coastal exposure to Atlantic variability, significant precursors only begin to emerge 4 days before rainfall events during DJF, and only 2 days before in JJA and MAM. 
In DJF, an initial sub-tropical ridge off the coast of Morocco develops into a meridional dipole, with low geopotential height over the North sea and a intensified central low-level jet 4 days prior to heavy rainfall. By lag 2 this anomaly strengthens, with high IVT anomalies coincident with the jet, indicating an influx of Atlantic moisture into the region. At the same time, upper-level wave activity becomes prominent, with some possible indications of upstream forcing from a weak wave-like geopotential anomaly over the US. By the day of the event itself, the V300 field shows a quadrapole structure supporting strong moisture convergence into the region. The low geopotential heights in the precursor pattern are more diffuse than for the other regions considered. As the domain is more isotropic, the exact placement of weather features over the domain is less stronglyy constrained, resulting in a greater `blurring' of the precursor pattern. Indeed, cyclones impacting this region are known to possess many disparate paths of evolution \citep{Hofstatter2018}. A qualitatively similar, but less robust series of precursors is visible for MAM and SON, with less large-scale conditioning.

Surprisingly, JJA rainfall events have quite extended and remote precursor patterns with a similar North American low appearing 4 days prior as was seen for SON West Iberian rainfall events. This is followed by the formation of a localised low over the UK, displaced South compared to other seasons with an accompanying upper-level wave visible at 2 days prior, possibly as a result of downstream development. There is no accompanying southerly high geopotential height as in other seasons indicating a less zonal flow and a more localised cut-off. The high pressure precursor over Italy at 1 day prior suggests a Mediterranean source of moisture is relevant, and capturing the footprint of the rare but impactful `Vb' cyclones \citep{Hofstatter2018}.

We have shown for three disparate European regions that our automated precursor detection approach is able to reproduce key insights from previous analyses of regional extreme precursors, and via its systematic application we have been able to discuss interseasonal variability fairly comprehensively, and in a way that enables intercomparison between regions. The purported value of this approach is that by conversion of precursor patterns into precursor activity indices we will be able to likewise convert these dynamical insights into improved event predictability. 

In the next section, we will validate this claim, zooming out from detailed consideration of a few case study regions to characterise the predictive power of precursor indices for all 18 European regions. Equivalent plots to figure \ref{fig:synth_plot} for all regions can be found in the supplementary material.

\section{European heavy rainfall precursors and prediction}\label{sec:pred}

\begin{figure}
    \centering
    \includegraphics[width=\linewidth]{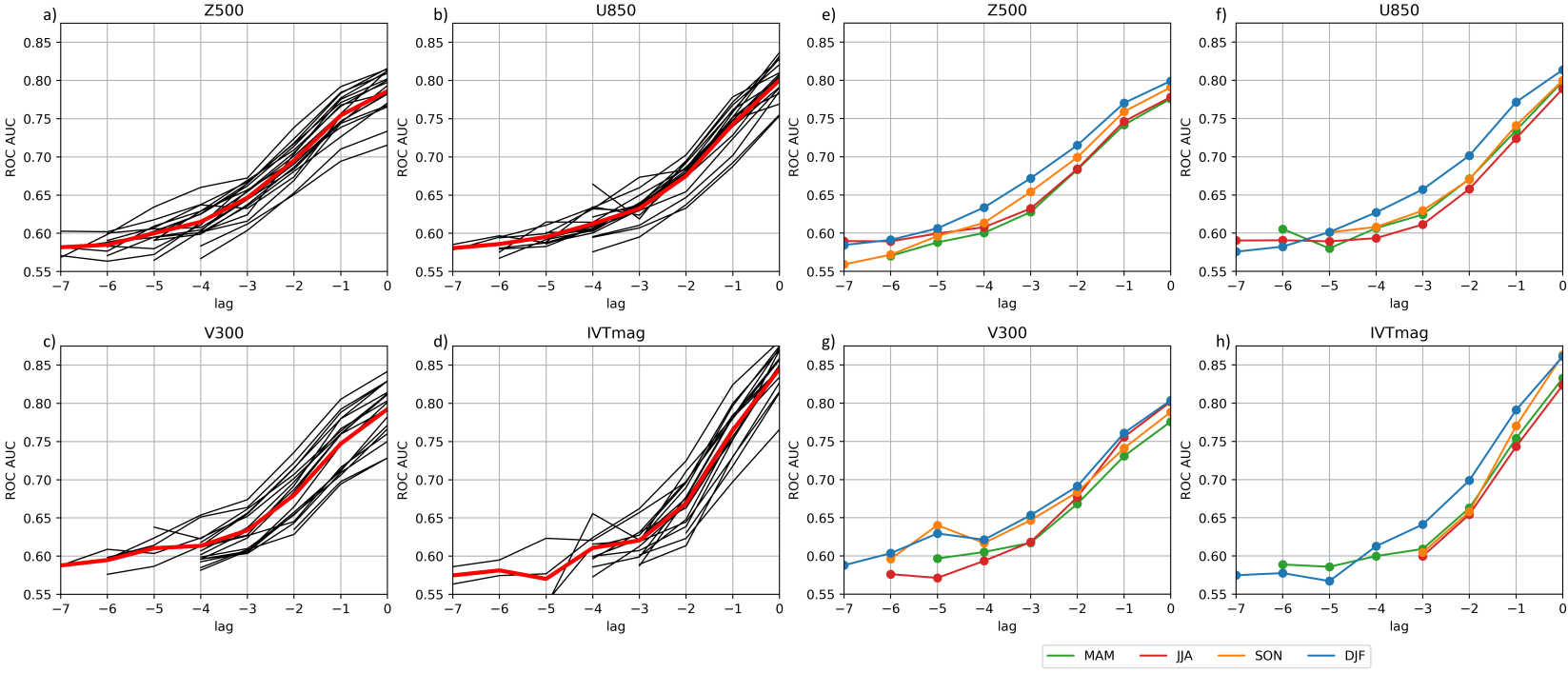}
    \caption{a) Predictive skill of precursor indices, as measured by cross-validated ROC AUC score averaged across seasons for each of 500hPa geopotential height (Z500), 850hPa zonal wind (U850), 300 hPa meridional wind (V300) and the magnitude of column-integrated vapour transport (IVTmag), with the skill for each region shown by a black line and the mean across both seasons and regions shown in red. b) ROC AUC now averaged across regions and conditioned on season.
    b) ROC AUC averaged across seasons}
    \label{fig:ROC}
\end{figure}

In this section we turn our attention to the precursor activity indices that correspond to each precursor pattern, and show that they have operational utility; that is, the potential to improve the predictability of heavy rainfall events. Without analysing forecast data, which will be the subject of future work, we estimate a lower-bound on predictive power by assessing the skill of logistic regression models fitted on reanalysis precursor indices for the prediction of heavy rainfall probability in each region, as discussed in section \ref{sec:methods_data}. Heuristically we consider a ROC AUC score >0.6 to indicate some useful amount of predictive power in a precursor.

By using a multivariate or nonlinear model, it is likely considerably higher skill could be reached, while in a hybrid dynamical-statistical prediction framework the predictive value will be closely tied to the forecast model's ability to identify precursors at extended lead times, as will be discussed in further detail in section \ref{sec:conclusions}. In fact, for the specific case of North Italy, a hybrid dynamical/machine-learning model leveraging these non-local precursors has already been found to boost skill out to day 10 \citep{Grazzini2023}.

For now, we proceed by analysing the univariate logistic-regression based ROC AUC of precursor indices in the European regions shown in figure \ref{fig:regions}.  Figure \ref{fig:ROC}a) shows the precursor skill averaged over season (for all seasons for which a precursor pattern is defined), and showing each region separately. As previous studies on extreme event precursors have focused on targeted regions and are often directed by specialist intuition that precursors should exist, it is by no means obvious that extremes over a generic region will or should have predictive long-range precursors. Through our systematic multi-region analysis, we are now able to comment on this point. There is substantial inter-regional spread in precursor skill; approximately double that of the seasonal spread. Differences are largest at short lead times, with precursor skill between regions converging at lags 3-5, and generally falling below 0.6 during this period.

It is interesting to consider the worst case scenarios, that is the regions with the lowest skill for each lead time and variable. The regions with the least skilful precursors nevertheless show ROC AUC above 0.7 for day 0 precursors, and maintain non-negligible skill for Z500, U850 and V300 out to lag 2. These least predictable regions are in all cases one of CZ and NE\_Europe (see SI figure 2 for a more detailed presentation), both of which are isolated from the upstream Atlantic flow. Many regions show much greater precursor skill and in the best cases there are indications that even lag 5 precursors have useful levels of skill. 

Summarising another way, figure \ref{fig:ROC}b) shows the ROC AUC for precursors based on each large-scale predictor variable, conditional on season and averaged across the 18 regions. For all variables, precursor skill is highest in DJF, generally followed by SON, with MAM and JJA showing similar, lower skill, with a total inter-seasonal spread of approximately 1 day of predictability. This corresponds to the known increased importance of large-scale forcing in DJF, and the increased role of convective processes in JJA. This skill difference may then be in part derive from choice of precursor variables, which emphasises the large-scale flow. While early experiments including CAPE and SST fields did not reveal skilful precursor patterns, it is possible a more detailed consideration of such thermodynamic variables might reveal better summertime precursors.

IVTmag precursors show the highest skill at short timescales but decay rapidly, with ROC AUC dropping below 0.6 at lags 3 and 4 for non-winter and wintertime precursors respectively. Short-lag Z500 precursors are comparatively less skilful but maintain predictive power out to lag 5 on average, which is consistent with the general large-scale wave-like precursors often seen for Z500, rather than the more isolated streaks of moisture flux seen in IVTmag. V300 and U850 precursors fall in between these two extremes, although U850 is notably more skilful in DJF, in keeping with the dominant role of DJF Atlantic trimodal jet variability \citep{Woollings2010}. 

For simplicity, we are working within a categorical event context here, both to identify our precursor indices, and to test their predictive skill. However it is interesting to consider whether the precursor indices are able to assess the amplitude of an extreme event. This is primarily left to future work but in supplementary figure 3 we show the value of precursor indices for the most extreme daily rainfall total in each region and season, according to ERA5. There are several notable cases where these extremes are associated with index values -- usually for IVTmag -- >6 standard deviations above the norm, suggesting that this is indeed the case. 

\begin{figure}
    \centering
    \includegraphics[width=\linewidth]{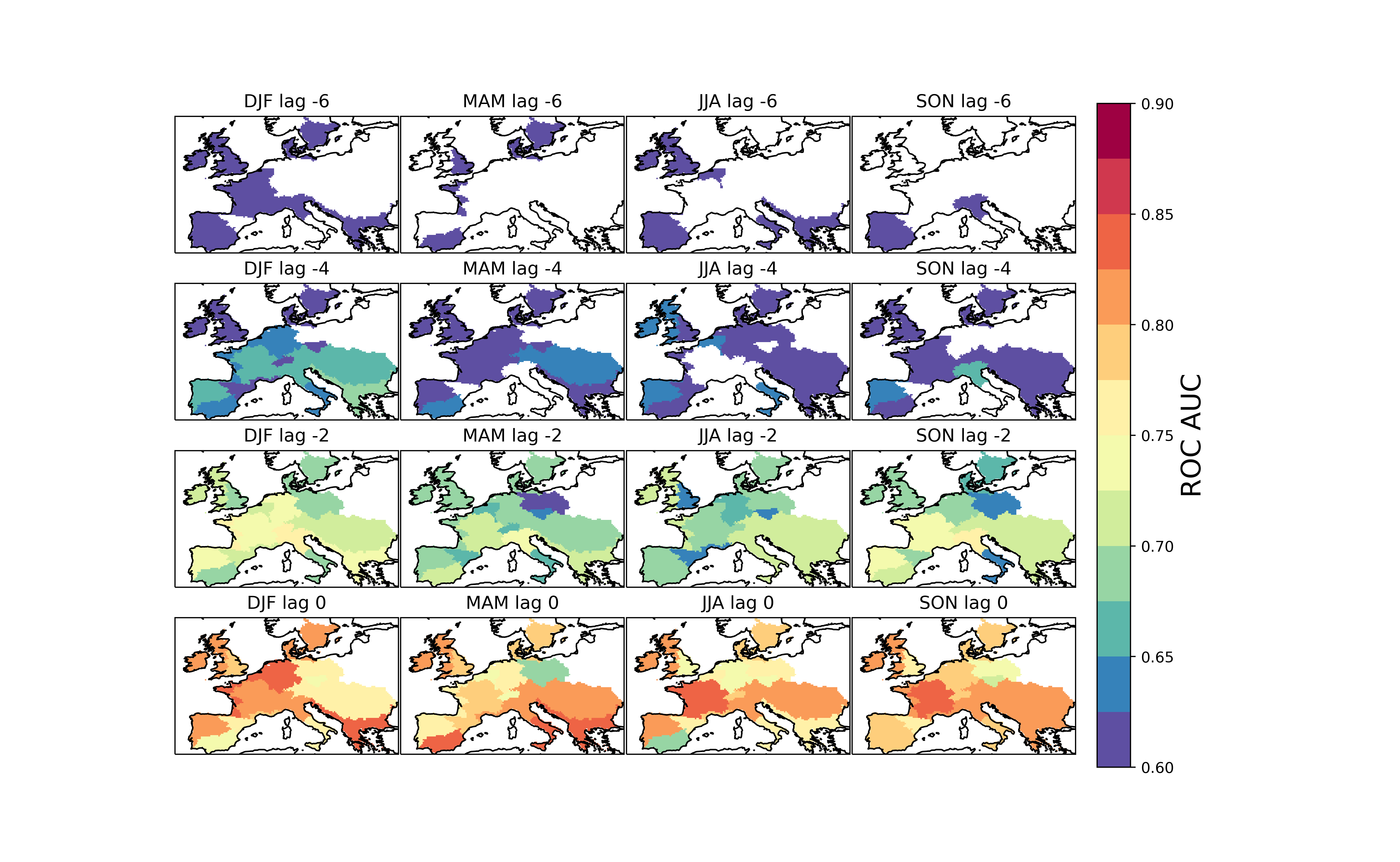}
    \caption{Z500 precursor ROC AUC for each spatial region, as a function of lag and season.}
    \label{fig:Z_skill}
\end{figure}

The role of spatial variability in shaping precursor skill can be seen more clearly in figure \ref{fig:Z_skill} for Z500 precursors. Equivalent skill maps for the other large-scale variables can be found in the supplementary data. In general, the Z500 precursor patterns considered represent either upstream wavetrains or blocking events at long lead times, and localised troughs at shorter lead times. We see that coastal regions show the longest range precursors, at lags of 6, both on the Atlantic coast, as for Iberian and UK regions, but also along the Eastern Mediterranean, where Italy and the Adriatic coast show early precursors especially in DJF. Most regions start to show precursors, often skilful ones, from day 4 onwards, often associated with the first indications of local trough formation. More mountainous domains tend to show higher precursor skill, a consequence of the greater constraints placed on the large-scale flow to allow substantial vapour transport into the domain. There is certainly no simple overall relation between location and precursor skill however: even neighbouring regions can have very different precursor skill and different seasonal variability. This motivates exactly the sort of spatially differentiated approach we have taken here.

\section{Synthesising precursors across regions}\label{sec:plsr}

In the previous sections we have demonstrated the ability of automated precursor detection to reveal the dynamical drivers, and aid the prediction, of extreme events across a large number of European domains. For many applications such a regional perspective is ideal, but there are also applications for which a broader perspective is useful. For example, operational centres responsible for an entire continent (such as ECWMF or NWS) might wish to monitor the risk of extreme events in geographically and administratively distinct regions of their domain, but without tracking individual precursors for all regions. Entirely equivalently, a national forecaster whose domain features great climactic diversity may wish to apply the precursor approach outlined above to sub-regions or individual stations, but then also to synthesise extreme event risk holistically. To address this use-case we now demonstrate how precursors for different regions can be synthesised down to a minimal set of indices that balance predictability in different regions, using PLSR regression.

\begin{figure}
    \centering
    \includegraphics[width=0.9\linewidth]{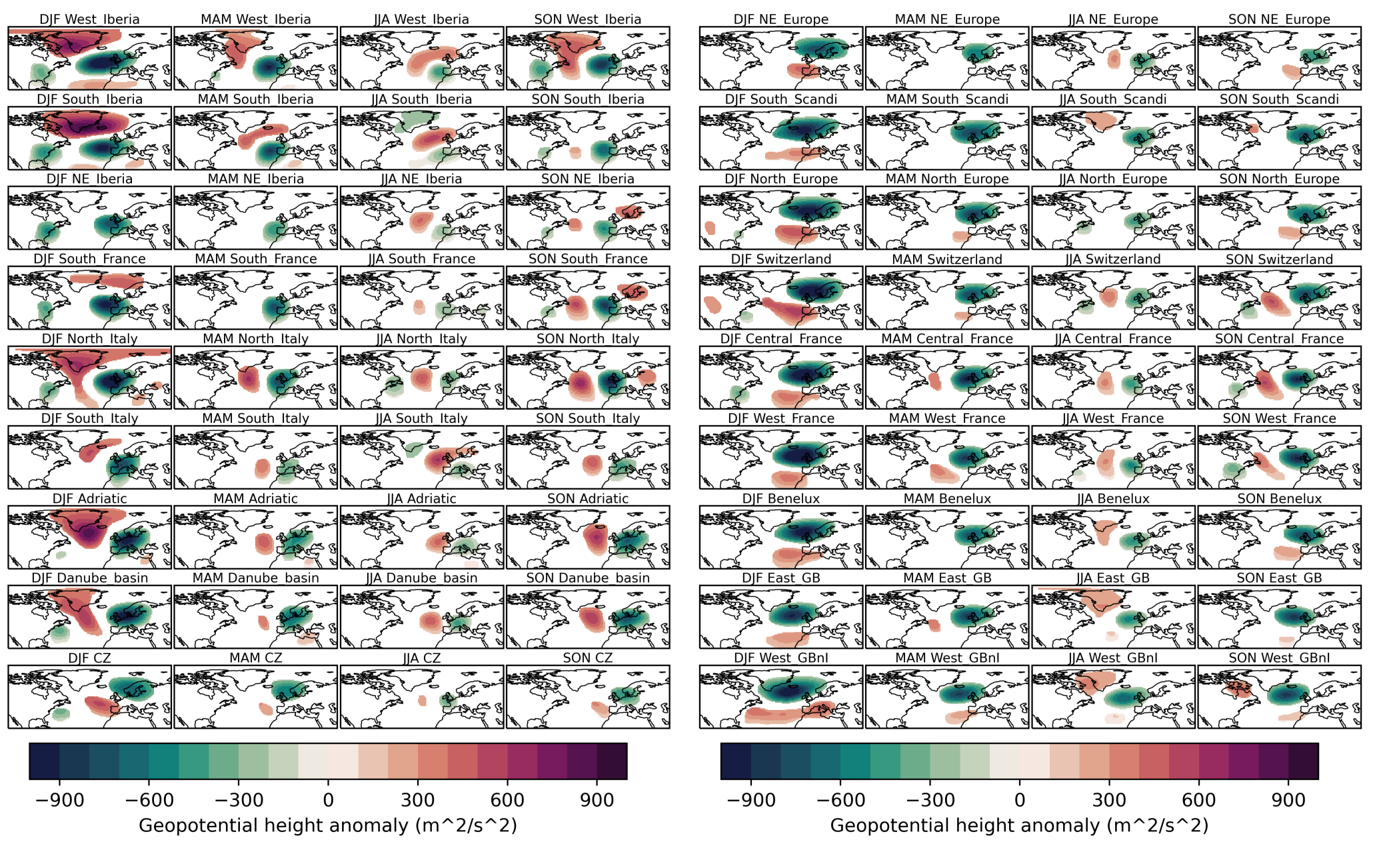}
    \caption{Lag 2 Z500 anomaly precursor patterns for each region and season. Clear similarities can be seen between many neighbouring regions.}
    \label{fig:stamp_plot}
\end{figure}

Our task is simplified in many cases by the fact that precursor patterns for extremes in neighbouring regions can often be very similar. This is demonstrated for lag 2 Z500 precursors in figure \ref{fig:stamp_plot}, which is representative of many precursor patterns, and which suggest that it should be possible to reduce the dimensionality of an initially large number of precursor activity indices. By doing so, we aim to obtain a few general modes relevant for many regions. Concretely, in our case we see that seasonal differences for particular regions are often very pronounced, while qualitative similarities between regional precursor patterns for a given season are easy to see in figure \ref{fig:stamp_plot}: the positive NAO-like pattern characteristic of DJF precursors in north-west domains and the negative NAO-like patterns in Mediterranean domains, the JJA south-easterly wave visible in a large number of regions, and the more zonally propagating wave that dominates SON central and southern Europe precursors with only slight phase differences. Further, as can be seen in figure \ref{fig:synth_plot}, the precursor patterns for sequential time lags can also be very similar.

\begin{figure}
    \centering
    \includegraphics[width=\linewidth]{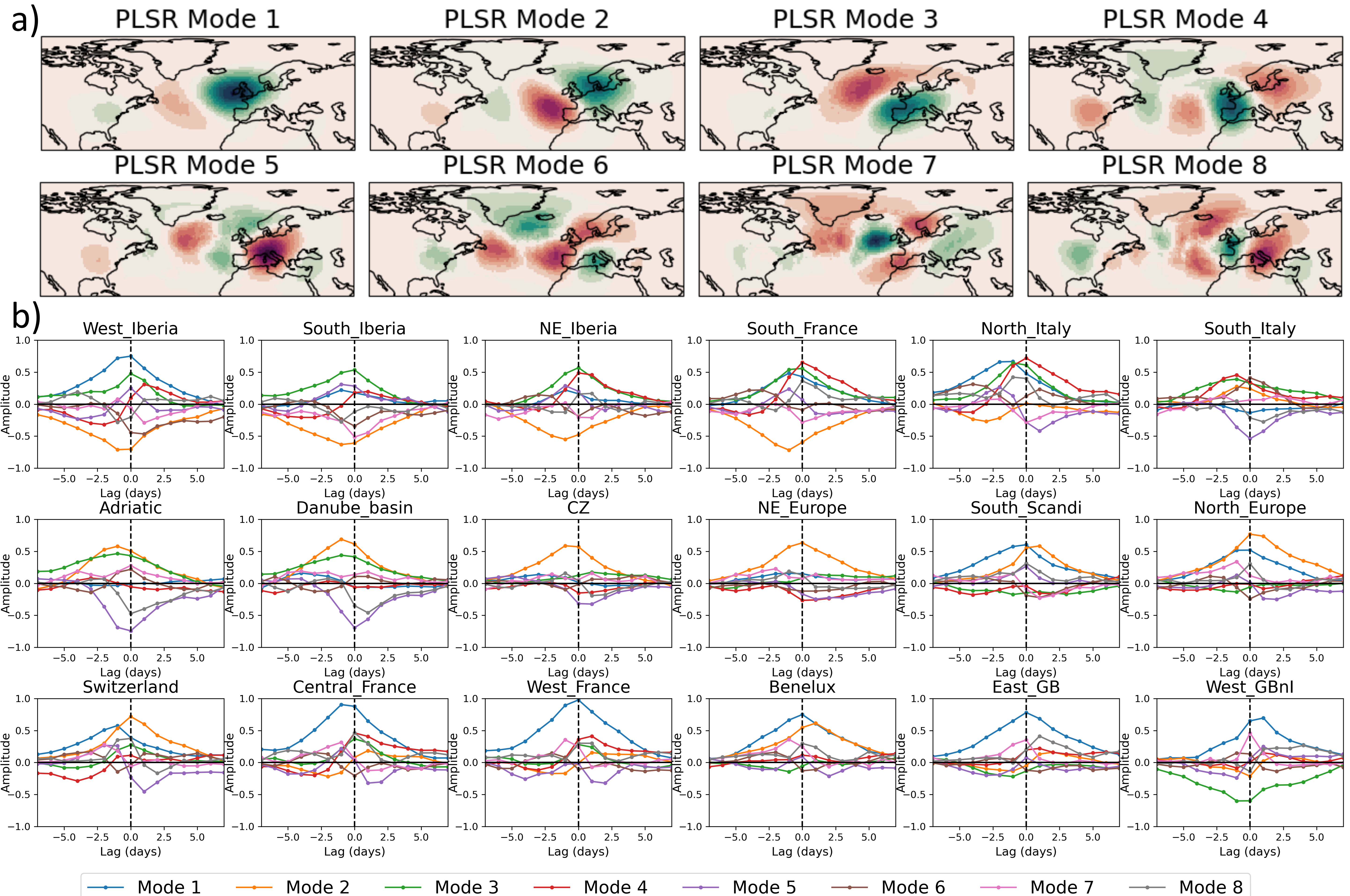}
    \caption{Z500 anomaly patterns corresponding to each PLSR mode of SON Z500 precursors across lags 5 to 0. b) The mean composite of each PLSR mode relative to occurrence of a SON heavy rainfall event in each region.}
    \label{fig:plsr_patterns}
\end{figure}

Without carrying out an exhaustive synthesis for all variables and seasons, we demonstrate how a minimal set of precursor indices can be produced for Z500 SON precursors, with a method that can be easily applied to any other season or variable of interest. Z500 precursors in SON generally show high predictive power at long lags, while large-scale SON dynamics are not as comprehensively studied as those of DJF, making this a natural combination to trial our approach on.

We apply partial-least-squares regression (PLSR) as discussed in section \ref{sec:methods}, searching for the $n$ orthogonal linear combinations (PLSR modes) of our predictor indices that explain the most variance in our event variables. As our predictors, we take SON Z500 precursor indices spanning from lags 5 to 0 for each of the 18 regions (108 variables total, reduced to 88 after discarding blank precursor patterns). For the target variables we use the heavy rainfall event index for each of the 18 regions, but as we seek modes that capture precursors at different lead times, i.e. we want to include lag-covariance, we duplicate each event index and offset it in time between 0 and 5 days, producing a corresponding 108-dimensional event index as our target variable.

Figure \ref{fig:plsr_patterns}a) shows Z500 anomaly patterns corresponding to the first eight PLSR modes. As for empirical orthogonal functions (EOFs), the sign of the anomaly patterns is arbitrary, each representing a symmetric mode of variability. Unlike EOFs of Z500 however, which tend to be dominated by large-scale dipole patterns, the resulting PLSR modes clearly project onto small scale flow features that are important for extreme European precipitation events. Larger scale trough/ridge patterns are visible in PLSR modes 1 and 3, zonal wave trains in modes 2, 4 and 5, and cutoff lows/highs in modes 7 and 8. 

In exchange for reducing 88 initial precursors to 8 modes, the interpretation of the PLSR modes' impact on rainfall becomes more complex: positive values of any given mode will increase heavy rainfall occurrence in some regions and at certain lead times over the following five days, and decrease it in others. Furthermore, as the PLSR indices are uncorrelated to eachother by construction, the covariability of the indices prior to an event carries additional information. This can be seen in figure \ref{fig:plsr_patterns}b) which shows the average value of each PLSR index in the days preceding and following SON heavy rainfall events in each region. For North Italy as an example, pronounced high or low values of almost all modes occur in the lead up to a heavy rainfall event: high modes 1 and 6 combining to approximate the lag 4 wavetrain precursor seen in figure \ref{fig:synth_plot}b), low values of mode 2 indicating the movement of the wave eastward by lag 2, and strong amplification of mode 4 as the wave reaches its final position just to the west of Italy.

\begin{figure}
    \centering
    \includegraphics[width=0.8\linewidth]{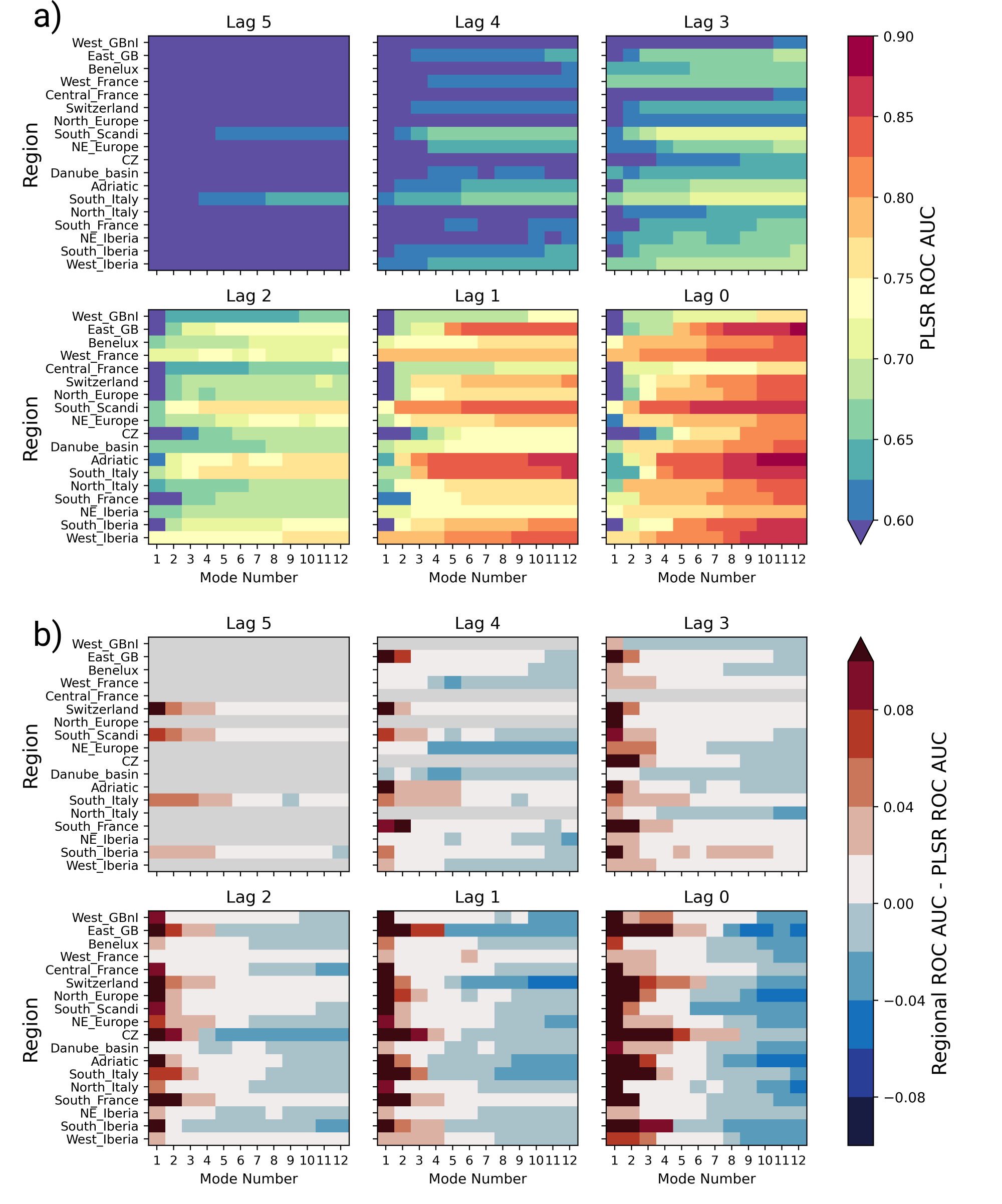}
    \caption{ROC AUC skill score for SON Z500 PLSR modes used to predict SON heavy rainfall via logistic regression (a) and the difference between PLSR skill and the corresponding SON Z500 precursor activity index skill (b) in each region as a function of region and number of modes used. In b), red values indicate regional indices are more predictive, while blue values indicate the PLSR modes are more predictive and lag-region combinations with no defined Z500 SON precursor pattern are greyed out.}
    \label{fig:plsr_skill}
\end{figure}

The decrease of straightforward interpretability that comes from PLSR reduction perhaps makes them most suitable for use in statistical prediction models rather than manual operational monitoring. Indeed we find that the predictive power of the PLSR indices compares favourably to the predictive power of the regionally-specific precursor indices considered previously. Figure \ref{fig:plsr_skill}a) shows the ROC AUC of statistical predictions of SON regional heavy rainfall using the same cross-validated logistic regression approach as in section \ref{sec:methods_skill} but now using different numbers of PLSR indices as predictive variables. The difference between the PLSR-based skill and the skill found for SON Z500 precursors for each region is shown in figure \ref{fig:plsr_skill}b). In general we find that the regional indices are more skilful than predictions based on <5 PLSR indices, with little improvement seen from using more than 8 PLSR indices. This is particularly the case when looking at lag 0 patterns, where the trough driving rainfall in each region is most localised and so least well approximated by general modes. However, overall we see that the degrees of freedom in our precursors can be reduced by a factor of 10, without reducing statistical skill. We therefore suggest PLSR reduction may be a useful post-processing step for multi-domain applications to simplify workflows and for deep learning approaches.

\section{Discussion and conclusion}\label{sec:conclusions}
The `quiet revolution' of steadily improving weather forecast skill has been achieved through the gradual synthesis and operational application of improved meteorological understanding \citep{Bauer2015}. In this paper we have presented a novel framework, termed `Domino', for identifying the large-scale precursor patterns of extreme weather events, and reducing those to data-efficient, predictive scalar activity indices. In doing so we provide a starting point for the operationalisation of research into the conditioning of extreme events on large-scale dynamics.

\subsection{Summary of results}\label{sec:conclusions_summary}

We have tested our `Domino' precursor identification framework by analysing regional daily heavy precipitation across western and central Europe, and across seasons. Even estimating predictive skill using only a simple univariate regression framework, we have shown that in all cases at least some degree of usable predictability (ROC AUC>0.6) is present in large-scale precursors based on geopotential height, wind fields and vapour transport patterns. This conditioning is in general lowest in JJA when meso-scale dynamics are more dominant, and in flatter, more continental domains. Even in these cases, predictive precursors are seen three days ahead, while in DJF, where conditioning on the large-scales is strongest, some domains indicate borderline-useable skill 6 days ahead.
We have demonstrated that the precursor patterns identified are consistent with the previously known dynamical precursors for three diverse regions, and by virtue of our systematic approach, are able to comprehensively characterise them.
Finally, we have demonstrated that many precursor patterns are similar between different domains; that is to say, the same large-scale dynamics favour rainfall in multiple regions. We show how this can be used to simplify the analytical framework without giving up regional specificity by synthesising the precursor indices for different regions together into a lower-dimensional set of predictive modes, using partial least squares regression. For the specific case of SON Z500 precursors, we show a ~10x reduction in dimensionality can be achieved with only minor decrease in predictive skill and a more complex physical interpretation, a trade-off that may be useful for certain applications. This synthesis is enabled by the use of precursor patterns to generate scalar precursor indices, and indicates the value of this novel approach. 

\subsection{Caveats}\label{sec:conclusions_caveats}

Naturally, there are limitations of the Domino framework which must be discussed.

Most obviously, as a bottom-up data-driven method the precursor patterns will be sensitive to changes in the event time series, with longer time series providing more robust results. The convolution filter applied during pattern computation counteracts this to an extent, by smoothing the pattern mask. In this paper we have used ~O(300) events for each region, which was found in early tests to provide qualitatively identical results for different data subsamples.

Secondly, as the framework is based on lagged composites of potentially predictive variables over all events, it is not possible to distinguish distinct dynamical pathways leading to the same event; instead the precursor pattern will contain a superposition of both. In the pathological case where two precursor pathways have opposite circulations then they would cancel, producing no significant composite. This can be mitigated by subsetting the extreme events if such pathways are suspected a priori (e.g. by splitting by season as done here), or by choice of variables (e.g. wave envelope amplitude vs V300 to prevent phase cancellation). Another option is to compare the variance of the composite to the climatological value, an analysis supported by Domino, but which has not been included in this work for simplicity. A related point is the implicit stationarity assumption at play. Non-stationary dynamics can essentially be understood as multiple pathways occurring preferentially during different time periods. Applying Domino to rolling time windows of the event time series could reveal such underlying non-stationarity. 

Finally, the motivating hypothesis of our approach is that the predictive precursors we identify have potential operational use. Considering the time-scale of our precursors (i.e. days 2-6 ahead) and the quality of modern NWP systems, it is unlikely a purely statistical prediction based on precursors will outperform dynamical forecasts in terms. Instead quantitative value may come from hybrid forecasting; using the dynamical model to predict the precursor patterns at extended range lead times, and then infer rainfall probability even further ahead from the precursors. For this to be relevant, it requires that at some lead time the dynamical model has skill at predicting the large-scale precursors, but no longer has skill at small-area rainfall. We consider this highly plausible, and \citet{Grazzini2023} has recently shown the assumption to hold for the North Italy region, but showing this is generally the case is an important question left for future work.

In addition, the operational analysis of precursors has narrative value for forecasters, helping to understand the range of possibilities and developing dynamical event precursors in a forecast, and so in general boosting their awareness of plausible weather beyond the limits of formalised skill. In a sense, analysing the unfurling preconditions to extreme events can serve to precondition the forecaster, allowing them to more quickly interpret and act on probable extreme events as they emerge at shorter lead times. 

\subsection{Extension}\label{sec:conclusions_extension}
In this section, we present ideas and methodological advice for extensions and generalisations of the work done in this paper to different events, regions and timescales. 

In this paper we have aimed to provide a relatively straightforward proof-of-concept application, and there are multiple refinements that could be made for real-world flood-prediction applications. Firstly, rain gauge data or a local high-resolution gridded rainfall dataset could be easily used in place of the ERA5 precipitation data used to define events here. Spatial averaging can be avoided entirely to provide extra information on local extremes: precursor indices could be computed for each rain gauge or grid-point independently, and then synthesised via the PLSR approach described above. For a consideration of true flood risk instead of simply high precipitation risk hydrological discharge data be used to define events. In this case, including land-surface precursor variables such as soil saturation and snow accumulation are likely to be of value.

Extending beyond the logistic regression used here, the scalar precursor indices are well-suited for input into more advanced statistical and machine-learning models, already providing a low-dimensional summary of the flow. While raw data sometimes proves just as effective as physically refined data for deep-learning applications, at the least, precursor analysis can assist variable selection by highlighting the nonlocal regions and timescales that are relevant for a particular event. In this case, it may be useful to lower the predictive threshold used to define a 'useful' precursor, in order to include more marginal or conditional sources of skill.

Relatedly, the predictability of events we report here should not be understood as inherent, but also as a result of the precursor variables we have chosen. For summer rainfall, predictive thermodynamic precursors may exist for some regions, based on variables such as CAPE, static stability, and T2m, although the predictability of such processes is inherently much more limited than the large-scale dynamics we focus on here.

Generalisation to different regions is trivial, simply requiring different event data. However relevant precursor variables will of course be quite different in tropical or sub-tropical regions where the thermodynamic variables mentioned above may contribute more significantly, and in strongly continental climates, where land-surface variable are likely more relevant. As a coastal, mid-latitude region, the variables considered here may extend most straightforwardly to the West coast US, perhaps also with the addition of Pacific SSTs.

It is similarly easy to move beyond rainfall extremes; wind extremes or extreme waves are both impactful on daily timescales, and could be treated in much the same way as rainfall. For longer timescale events such as heatwaves, or even relevant non-extremes such as predicting monthly rainfall terciles, the main methodological differences come as a result of reduced sample size. When computing the pattern mask, a reduced amplitude mask and increased convolution filter may serve to increase sensitivity to weak sources of S2S predictability while helping to maintain robust patterns. Variables such as sea-ice extent and SSTs, and stratospheric wind are natural candidates for such a study.  

Finally, beyond prediction, automated precursor analysis has value both for the fundamental study of dynamics and for model validation purposes. Just as circulation regimes are used to explore the dynamics of surface teleconnections \citep{Domeisen2020}, sources of model bias \citep{Wandel2023} and to judge the performance of climate models \citep{Dorrington2022}, so event-centric precursor patterns and indices could be used to summarise the representation of different dynamical processes in models. Vice-versa, events could be defined in terms of dynamical features -- such as sudden stratospheric warmings, regime transitions or wave breaking events -- and their precursor and `postcursor' indices can be used to help understand the variability between events by, for example, clustering in the space of indices.

In our own continuing work, we aim to demonstrate how precursors can be used to summarise and digest the behaviour of large forecast ensembles, and to develop an operational early-warning system for extreme events.

\section*{scripts and datasets}
ERA5 data used in this study is freely available from the Copernicus Data Store. Software tools required to duplicate or extend this work are available as part of the Domino package: \verb|https://github.com/joshdorrington/domino|. A compressed folder of further supplementary figures is available at \verb|https://doi.org/10.5281/zenodo.8087482|

\bibliography{domino2023v3}

\begin{thebibliography}{58}
\expandafter\ifx\csname natexlab\endcsname\relax\def\natexlab#1{#1}\fi
\expandafter\ifx\csname url\endcsname\relax
  \def\url#1{\texttt{#1}}\fi
\expandafter\ifx\csname urlprefix\endcsname\relax\def\urlprefix{URL: }\fi

\bibitem[{skl(2023)}]{sklearn-plsr}
 (2023) Sklearn.cross\_decomposition.{{PLSRegression}}.
\newblock
  https://scikit-learn/stable/modules/generated/sklearn.cross\_decomposition.PLSRegression.html.

\bibitem[{Abdi(2010)}]{Abdi2010}
Abdi, H. (2010) Partial least squares regression and projection on latent
  structure regression ({{PLS Regression}}).
\newblock \textit{WIREs Computational Statistics}, \textbf{2}, 97--106.

\bibitem[{AghaKouchak et~al.(2022)AghaKouchak, Pan, Mazdiyasni, Sadegh, Jiwa,
  Zhang, Love, Madadgar, Papalexiou, Davis, Hsu and
  Sorooshian}]{AghaKouchak2022}
AghaKouchak, A., Pan, B., Mazdiyasni, O., Sadegh, M., Jiwa, S., Zhang, W.,
  Love, C.~A., Madadgar, S., Papalexiou, S.~M., Davis, S.~J., Hsu, K. and
  Sorooshian, S. (2022) Status and prospects for drought forecasting:
  Opportunities in artificial intelligence and hybrid physicalstatistical
  forecasting.
\newblock \textit{Philosophical Transactions of the Royal Society A}.

\bibitem[{Barton et~al.(2016)Barton, Giannakaki, {von Waldow}, Chevalier, Pfahl
  and Martius}]{Barton2016}
Barton, Y., Giannakaki, P., {von Waldow}, H., Chevalier, C., Pfahl, S. and
  Martius, O. (2016) Clustering of {{Regional-Scale Extreme Precipitation
  Events}} in {{Southern Switzerland}}.
\newblock \textit{Monthly Weather Review}, \textbf{144}, 347--369.

\bibitem[{Barton et~al.(2022)Barton, Rivoire, Koh, Ali~S., Kopp and
  Martius}]{Barton2022}
Barton, Y., Rivoire, P., Koh, J., Ali~S., M., Kopp, J. and Martius, O. (2022)
  On the temporal clustering of {{European}} extreme precipitation events and
  its relationship to persistent and transient large-scale atmospheric drivers.
\newblock \textit{Weather and Climate Extremes}, \textbf{38}, 100518.

\bibitem[{Bauer et~al.(2015)Bauer, Thorpe and Brunet}]{Bauer2015}
Bauer, P., Thorpe, A. and Brunet, G. (2015) The quiet revolution of numerical
  weather prediction.
\newblock \textit{Nature}, \textbf{525}, 47--55.

\bibitem[{Beerli and Grams(2019)}]{Beerli2019}
Beerli, R. and Grams, C.~M. (2019) Stratospheric modulation of the large-scale
  circulation in the {{Atlantic}}\textendash{{European}} region and its
  implications for surface weather events.
\newblock \textit{Quarterly Journal of the Royal Meteorological Society},
  \textbf{145}, 3732--3750.

\bibitem[{Bloomfield et~al.(2021)Bloomfield, Brayshaw, Gonzalez and
  {Charlton-Perez}}]{Bloomfield2021}
Bloomfield, H.~C., Brayshaw, D.~J., Gonzalez, P. L.~M. and {Charlton-Perez}, A.
  (2021) Pattern-based conditioning enhances sub-seasonal prediction skill of
  {{European}} national energy variables.
\newblock \textit{Meteorological Applications}, \textbf{28}, e2018.

\bibitem[{Domeisen et~al.(2020)Domeisen, Grams and Papritz}]{Domeisen2020}
Domeisen, D. I.~V., Grams, C.~M. and Papritz, L. (2020) The role of {{North
  Atlantic}}\textendash{{European}} weather regimes in the surface impact of
  sudden stratospheric warming events.
\newblock \textit{Weather and Climate Dynamics}, \textbf{1}, 373--388.

\bibitem[{Dorrington et~al.(2022)Dorrington, Strommen and
  Fabiano}]{Dorrington2022}
Dorrington, J., Strommen, K. and Fabiano, F. (2022) Quantifying climate model
  representation of the wintertime {{Euro-Atlantic}} circulation using
  geopotential-jet regimes.
\newblock \textit{Weather and Climate Dynamics}, \textbf{3}, 505--533.

\bibitem[{EEA(2022)}]{EEA2022}
EEA (2022) Economic losses and fatalities from weather- and climate-related
  events in {{Europe}} \textemdash{} {{European Environment Agency}}.
\newblock
  https://www.eea.europa.eu/publications/economic-losses-and-fatalities-from.

\bibitem[{{Eiras-Barca} et~al.(2016){Eiras-Barca}, Brands and
  {Miguez-Macho}}]{Eiras-Barca2016}
{Eiras-Barca}, J., Brands, S. and {Miguez-Macho}, G. (2016) Seasonal variations
  in {{North Atlantic}} atmospheric river activity and associations with
  anomalous precipitation over the {{Iberian Atlantic Margin}}.
\newblock \textit{Journal of Geophysical Research: Atmospheres}, \textbf{121},
  931--948.

\bibitem[{{Eiras-Barca} et~al.(2018){Eiras-Barca}, Lorenzo, Taboada, Robles and
  {Miguez-Macho}}]{Eiras-Barca2018}
{Eiras-Barca}, J., Lorenzo, N., Taboada, J., Robles, A. and {Miguez-Macho}, G.
  (2018) On the relationship between atmospheric rivers, weather types and
  floods in {{Galicia}} ({{NW Spain}}).
\newblock \textit{Natural Hazards and Earth System Sciences}, \textbf{18},
  1633--1645.

\bibitem[{Ferranti et~al.(2015)Ferranti, Corti and Janousek}]{Ferranti2015}
Ferranti, L., Corti, S. and Janousek, M. (2015) Flow-dependent verification of
  the {{ECMWF}} ensemble over the {{Euro-Atlantic}} sector.
\newblock \textit{Quarterly Journal of the Royal Meteorological Society},
  \textbf{141}, 916--924.

\bibitem[{Fredriksen et~al.(2020)Fredriksen, Berner, Subramanian and
  Capotondi}]{Fredriksen2020}
Fredriksen, H.-B., Berner, J., Subramanian, A.~C. and Capotondi, A. (2020) How
  {{Does El Ni\~no}}\textendash{{Southern Oscillation Change Under Global
  Warming}}\textemdash{{A First Look}} at {{CMIP6}}.
\newblock \textit{Geophysical Research Letters}, \textbf{47}, e2020GL090640.

\bibitem[{Gasc{\'o}n et~al.(2023)Gasc{\'o}n, Montani and Hewson}]{Gascon2023}
Gasc{\'o}n, E., Montani, A. and Hewson, T.~D. (2023) Post-processing output
  from ensembles with and without parametrised convection, to create accurate,
  blended, high-fidelity rainfall forecasts.

\bibitem[{{Gonz{\'a}lez-Alem{\'a}n} et~al.(2022){Gonz{\'a}lez-Alem{\'a}n},
  Grams, Ayarzag{\"u}ena, {Zurita-Gotor}, Domeisen, G{\'o}mara,
  {Rodr{\'i}guez-Fonseca} and Vitart}]{Gonzalez-Aleman2022}
{Gonz{\'a}lez-Alem{\'a}n}, J.~J., Grams, C.~M., Ayarzag{\"u}ena, B.,
  {Zurita-Gotor}, P., Domeisen, D. I.~V., G{\'o}mara, I.,
  {Rodr{\'i}guez-Fonseca}, B. and Vitart, F. (2022) Tropospheric {{Role}} in
  the {{Predictability}} of the {{Surface Impact}} of the 2018 {{Sudden
  Stratospheric Warming Event}}.
\newblock \textit{Geophysical Research Letters}, \textbf{49}, e2021GL095464.

\bibitem[{Grams et~al.(2017)Grams, Beerli, Pfenninger, Staffell and
  Wernli}]{Grams2017}
Grams, C.~M., Beerli, R., Pfenninger, S., Staffell, I. and Wernli, H. (2017)
  Balancing {{Europe}}'s wind-power output through spatial deployment informed
  by weather regimes.
\newblock \textit{Nature Climate Change}, \textbf{7}, 557--562.

\bibitem[{Grams et~al.(2011)Grams, Wernli, B{\"o}ttcher, {\v C}ampa, Corsmeier,
  Jones, Keller, Lenz and Wiegand}]{Grams2011}
Grams, C.~M., Wernli, H., B{\"o}ttcher, M., {\v C}ampa, J., Corsmeier, U.,
  Jones, S.~C., Keller, J.~H., Lenz, C.-J. and Wiegand, L. (2011) The key role
  of diabatic processes in modifying the upper-tropospheric wave guide: A
  {{North Atlantic}} case-study.
\newblock \textit{Quarterly Journal of the Royal Meteorological Society},
  \textbf{137}, 2174--2193.

\bibitem[{Grazzini et~al.(2020{\natexlab{a}})Grazzini, Craig, Keil, Antolini
  and Pavan}]{Grazzini2020}
Grazzini, F., Craig, G.~C., Keil, C., Antolini, G. and Pavan, V.
  (2020{\natexlab{a}}) Extreme precipitation events over northern {{Italy}}.
  {{Part I}}: {{A}} systematic classification with machine-learning techniques.
\newblock \textit{Quarterly Journal of the Royal Meteorological Society},
  \textbf{146}, 69--85.

\bibitem[{Grazzini et~al.(2023)Grazzini, Dorrington, Grams, Craig, Magnusson
  and Vitart}]{Grazzini2023}
Grazzini, F., Dorrington, J., Grams, C.~M., Craig, G.~C., Magnusson, L. and
  Vitart, F. (2023) Boosting forecast of precipitation extremes over
  {{N-Italy}} using machine learning.
\newblock \textit{In Preparation}.

\bibitem[{Grazzini et~al.(2020{\natexlab{b}})Grazzini, Fragkoulidis, Pavan and
  Antolini}]{Grazzini2020a}
Grazzini, F., Fragkoulidis, G., Pavan, V. and Antolini, G. (2020{\natexlab{b}})
  The 1994 {{Piedmont}} flood: An archetype of extreme precipitation events in
  {{Northern Italy}}.
\newblock \textit{Bulletin of Atmospheric Science and Technology}, \textbf{1},
  283--295.

\bibitem[{Grazzini and Vitart(2015)}]{Grazzini2015}
Grazzini, F. and Vitart, F. (2015) Atmospheric predictability and {{Rossby}}
  wave packets.
\newblock \textit{Quarterly Journal of the Royal Meteorological Society},
  \textbf{141}, 2793--2802.

\bibitem[{Hanley and McNeil(1982)}]{Hanley1982}
Hanley, J.~A. and McNeil, B.~J. (1982) The meaning and use of the area under a
  receiver operating characteristic ({{ROC}}) curve.
\newblock \textit{Radiology}, \textbf{143}, 29--36.

\bibitem[{Hersbach et~al.(2020)Hersbach, Bell, Berrisford, Hirahara,
  Hor{\'a}nyi, {Mu{\~n}oz-Sabater}, Nicolas, Peubey, Radu, Schepers, Simmons,
  Soci, Abdalla, Abellan, Balsamo, Bechtold, Biavati, Bidlot, Bonavita,
  De~Chiara, Dahlgren, Dee, Diamantakis, Dragani, Flemming, Forbes, Fuentes,
  Geer, Haimberger, Healy, Hogan, H{\'o}lm, Janiskov{\'a}, Keeley, Laloyaux,
  Lopez, Lupu, Radnoti, {de Rosnay}, Rozum, Vamborg, Villaume and
  Th{\'e}paut}]{Hersbach2020}
Hersbach, H., Bell, B., Berrisford, P., Hirahara, S., Hor{\'a}nyi, A.,
  {Mu{\~n}oz-Sabater}, J., Nicolas, J., Peubey, C., Radu, R., Schepers, D.,
  Simmons, A., Soci, C., Abdalla, S., Abellan, X., Balsamo, G., Bechtold, P.,
  Biavati, G., Bidlot, J., Bonavita, M., De~Chiara, G., Dahlgren, P., Dee, D.,
  Diamantakis, M., Dragani, R., Flemming, J., Forbes, R., Fuentes, M., Geer,
  A., Haimberger, L., Healy, S., Hogan, R.~J., H{\'o}lm, E., Janiskov{\'a}, M.,
  Keeley, S., Laloyaux, P., Lopez, P., Lupu, C., Radnoti, G., {de Rosnay}, P.,
  Rozum, I., Vamborg, F., Villaume, S. and Th{\'e}paut, J.-N. (2020) The
  {{ERA5}} global reanalysis.
\newblock \textit{Quarterly Journal of the Royal Meteorological Society},
  \textbf{146}, 1999--2049.

\bibitem[{Hofst{\"a}tter et~al.(2018)Hofst{\"a}tter, Lexer, Homann and
  Bl{\"o}schl}]{Hofstatter2018}
Hofst{\"a}tter, M., Lexer, A., Homann, M. and Bl{\"o}schl, G. (2018)
  Large-scale heavy precipitation over central {{Europe}} and the role of
  atmospheric cyclone track types.
\newblock \textit{International Journal of Climatology}, \textbf{38},
  e497--e517.

\bibitem[{Holm(1979)}]{Holm1979}
Holm, S. (1979) A {{Simple Sequentially Rejective Multiple Test Procedure}}.
\newblock \textit{Scandinavian Journal of Statistics}, \textbf{6}, 65--70.

\bibitem[{Hoyer and Hamman(2017)}]{Hoyer2017}
Hoyer, S. and Hamman, J. (2017) Xarray: {{N-D}} labeled {{Arrays}} and
  {{Datasets}} in {{Python}}.
\newblock \textit{Journal of Open Research Software}, \textbf{5}, 10.

\bibitem[{Hunt and Watkiss(2011)}]{Hunt2011}
Hunt, A. and Watkiss, P. (2011) Climate change impacts and adaptation in
  cities: A review of the literature.
\newblock \textit{Climatic Change}, \textbf{104}, 13--49.

\bibitem[{Leeding et~al.(2023)Leeding, Riboldi and Messori}]{Leeding2023}
Leeding, R., Riboldi, J. and Messori, G. (2023) On {{Pan-Atlantic}} cold, wet
  and windy compound extremes.
\newblock \textit{Weather and Climate Extremes}, \textbf{39}, 100524.

\bibitem[{Lehner and Grill(2013)}]{Lehner2013}
Lehner, B. and Grill, G. (2013) Global river hydrography and network routing:
  Baseline data and new approaches to study the world's large river systems.
\newblock \textit{Hydrological Processes}, \textbf{27}, 2171--2186.

\bibitem[{Leon(2023)}]{Leon2023}
Leon, J. A.~P. (2023) Scale-dependent verification of precipitation and
  cloudiness at {{ECMWF}}.
\newblock
  https://www.ecmwf.int/en/newsletter/174/earth-system-science/scale-dependent-verification-precipitation-and-cloudiness.

\bibitem[{Lorenz(1969)}]{Lorenz1969}
Lorenz, E.~N. (1969) The predictability of a flow which possesses many scales
  of motion.
\newblock \textit{Tellus}, \textbf{21}, 289--307.

\bibitem[{Maldonado et~al.(2013)Maldonado, Alfaro, {Fallas-L{\'o}pez} and
  Alvarado}]{Maldonado2013}
Maldonado, T., Alfaro, E., {Fallas-L{\'o}pez}, B. and Alvarado, L. (2013)
  Seasonal prediction of extreme precipitation events and frequency of rainy
  days over {{Costa Rica}}, {{Central America}}, using {{Canonical Correlation
  Analysis}}.
\newblock \textit{Advances in Geosciences}, \textbf{33}, 41--52.

\bibitem[{Martius et~al.(2008)Martius, Schwierz and Davies}]{Martius2008}
Martius, O., Schwierz, C. and Davies, H.~C. (2008) {Far-upstream precursors of
  heavy precipitation events on the Alpine south-side}.
\newblock \textit{Quarterly Journal of the Royal Meteorological Society},
  \textbf{134}, 417--428.

\bibitem[{Masato et~al.(2012)Masato, Hoskins and Woollings}]{Masato2012}
Masato, G., Hoskins, B.~J. and Woollings, T.~J. (2012) Wave-breaking
  characteristics of midlatitude blocking.
\newblock \textit{Quarterly Journal of the Royal Meteorological Society},
  \textbf{138}, 1285--1296.

\bibitem[{Mastrantonas et~al.(2021)Mastrantonas, {Herrera-Lormendez},
  Magnusson, Pappenberger and Matschullat}]{Mastrantonas2021}
Mastrantonas, N., {Herrera-Lormendez}, P., Magnusson, L., Pappenberger, F. and
  Matschullat, J. (2021) Extreme precipitation events in the {{Mediterranean}}:
  {{Spatiotemporal}} characteristics and connection to large-scale atmospheric
  flow patterns.
\newblock \textit{International Journal of Climatology}, \textbf{41},
  2710--2728.

\bibitem[{Messori et~al.(2016)Messori, Caballero and Gaetani}]{Messori2016}
Messori, G., Caballero, R. and Gaetani, M. (2016) On cold spells in {{North
  America}} and storminess in western {{Europe}}.
\newblock \textit{Geophysical Research Letters}, \textbf{43}, 6620--6628.

\bibitem[{Mora and Vieira(2020)}]{Mora2020}
Mora, C. and Vieira, G. (2020) The {{Climate}} of {{Portugal}}.
\newblock In \textit{Landscapes and {{Landforms}} of {{Portugal}}} (eds.
  G.~Vieira, J.~L. Z{\^e}zere and C.~Mora), World {{Geomorphological
  Landscapes}}, 33--46. {Cham}: {Springer International Publishing}.

\bibitem[{Oertel et~al.(2023)Oertel, Pickl, Quinting, Hauser, Wandel,
  Magnusson, Balmaseda, Vitart and Grams}]{Oertel2023}
Oertel, A., Pickl, M., Quinting, J.~F., Hauser, S., Wandel, J., Magnusson, L.,
  Balmaseda, M., Vitart, F. and Grams, C.~M. (2023) Everything {{Hits}} at
  {{Once}}: {{How Remote Rainfall Matters}} for the {{Prediction}} of the 2021
  {{North American Heat Wave}}.
\newblock \textit{Geophysical Research Letters}, \textbf{50}, e2022GL100958.

\bibitem[{Parker et~al.(2019)Parker, Woollings, Weisheimer, O'Reilly, Baker and
  Shaffrey}]{Parker2019}
Parker, T., Woollings, T., Weisheimer, A., O'Reilly, C., Baker, L. and
  Shaffrey, L. (2019) Seasonal {{Predictability}} of the {{Winter North
  Atlantic Oscillation From}} a {{Jet Stream Perspective}}.
\newblock \textit{Geophysical Research Letters}, \textbf{46}, 10159--10167.

\bibitem[{Pasquier et~al.(2019)Pasquier, Pfahl and Grams}]{Pasquier2019}
Pasquier, J.~T., Pfahl, S. and Grams, C.~M. (2019) Modulation of {{Atmospheric
  River Occurrence}} and {{Associated Precipitation Extremes}} in the {{North
  Atlantic Region}} by {{European Weather Regimes}}.
\newblock \textit{Geophysical Research Letters}, \textbf{46}, 1014--1023.

\bibitem[{Pedregosa et~al.(2011)Pedregosa, Varoquaux, Gramfort, Michel,
  Thirion, Grisel, Blondel, Prettenhofer, Weiss, Dubourg, Vanderplas, Passos
  and Cournapeau}]{Pedregosa2011}
Pedregosa, F., Varoquaux, G., Gramfort, A., Michel, V., Thirion, B., Grisel,
  O., Blondel, M., Prettenhofer, P., Weiss, R., Dubourg, V., Vanderplas, J.,
  Passos, A. and Cournapeau, D. (2011) Scikit-learn: {{Machine Learning}} in
  {{Python}}.
\newblock \textit{MACHINE LEARNING IN PYTHON}.

\bibitem[{Pino et~al.(2016)Pino, {Ruiz-Bellet}, Balasch, {Romero-Le{\'o}n},
  Tuset, Barriendos, Mazon and Castelltort}]{Pino2016}
Pino, D., {Ruiz-Bellet}, J.~L., Balasch, J.~C., {Romero-Le{\'o}n}, L., Tuset,
  J., Barriendos, M., Mazon, J. and Castelltort, X. (2016) Meteorological and
  hydrological analysis of major floods in {{NE Iberian Peninsula}}.
\newblock \textit{Journal of Hydrology}, \textbf{541}, 63--89.

\bibitem[{Ramos et~al.(2018)Ramos, Martins, Tom{\'e} and Trigo}]{Ramos2018}
Ramos, A.~M., Martins, M.~J., Tom{\'e}, R. and Trigo, R.~M. (2018) Extreme
  {{Precipitation Events}} in {{Summer}} in the {{Iberian Peninsula}} and {{Its
  Relationship With Atmospheric Rivers}}.
\newblock \textit{Frontiers in Earth Science}, \textbf{6}.

\bibitem[{Ramos et~al.(2015)Ramos, Trigo, Liberato and Tom{\'e}}]{Ramos2015}
Ramos, A.~M., Trigo, R.~M., Liberato, M. L.~R. and Tom{\'e}, R. (2015) Daily
  {{Precipitation Extreme Events}} in the {{Iberian Peninsula}} and {{Its
  Association}} with {{Atmospheric Rivers}}.
\newblock \textit{Journal of Hydrometeorology}, \textbf{16}, 579--597.

\bibitem[{{Raveh-Rubin} and Flaounas(2017)}]{Raveh-Rubin2017}
{Raveh-Rubin}, S. and Flaounas, E. (2017) A dynamical link between deep
  {{Atlantic}} extratropical cyclones and intense {{Mediterranean}} cyclones.
\newblock \textit{Atmospheric Science Letters}, \textbf{18}, 215--221.

\bibitem[{Richardson et~al.(2020)Richardson, Cloke and
  Pappenberger}]{Richardson2020}
Richardson, D.~S., Cloke, H.~L. and Pappenberger, F. (2020) Evaluation of the
  {{Consistency}} of {{ECMWF Ensemble Forecasts}}.
\newblock \textit{Geophysical Research Letters}, \textbf{47}, e2020GL087934.

\bibitem[{Rocklin(2015)}]{Rocklin2015}
Rocklin, M. (2015) Dask: {{Parallel Computation}} with {{Blocked}} algorithms
  and {{Task Scheduling}}.
\newblock In \textit{Python in {{Science Conference}}}, 126--132. {Austin,
  Texas}.

\bibitem[{Salgueiro et~al.(2013)Salgueiro, Machado, Barriendos, Pereira and
  Benito}]{Salgueiro2013}
Salgueiro, A.~R., Machado, M.~J., Barriendos, M., Pereira, H.~G. and Benito, G.
  (2013) Flood magnitudes in the {{Tagus River}} ({{Iberian Peninsula}}) and
  its stochastic relationship with daily {{North Atlantic Oscillation}} since
  mid-19th {{Century}}.
\newblock \textit{Journal of Hydrology}, \textbf{502}, 191--201.

\bibitem[{Santos et~al.(2018)Santos, Fragoso and Santos}]{Santos2018}
Santos, M., Fragoso, M. and Santos, J.~A. (2018) Damaging flood severity
  assessment in {{Northern Portugal}} over more than 150~years (1865\textendash
  2016).
\newblock \textit{Natural Hazards}, \textbf{91}, 983--1002.

\bibitem[{Shepherd et~al.(2018)Shepherd, Boyd, Calel, Chapman, Dessai,
  {Dima-West}, Fowler, James, Maraun, Martius, Senior, Sobel, Stainforth, Tett,
  Trenberth, {van den Hurk}, Watkins, Wilby and Zenghelis}]{Shepherd2018}
Shepherd, T.~G., Boyd, E., Calel, R.~A., Chapman, S.~C., Dessai, S.,
  {Dima-West}, I.~M., Fowler, H.~J., James, R., Maraun, D., Martius, O.,
  Senior, C.~A., Sobel, A.~H., Stainforth, D.~A., Tett, S. F.~B., Trenberth,
  K.~E., {van den Hurk}, B. J. J.~M., Watkins, N.~W., Wilby, R.~L. and
  Zenghelis, D.~A. (2018) Storylines: An alternative approach to representing
  uncertainty in physical aspects of climate change.
\newblock \textit{Climatic Change}, \textbf{151}, 555--571.

\bibitem[{Simpson et~al.(2020)Simpson, Bacmeister, Neale, Hannay, Gettelman,
  Garcia, Lauritzen, Marsh, Mills, Medeiros and Richter}]{Simpson2020}
Simpson, I.~R., Bacmeister, J., Neale, R.~B., Hannay, C., Gettelman, A.,
  Garcia, R.~R., Lauritzen, P.~H., Marsh, D.~R., Mills, M.~J., Medeiros, B. and
  Richter, J.~H. (2020) An {{Evaluation}} of the {{Large-Scale Atmospheric
  Circulation}} and {{Its Variability}} in {{CESM2}} and {{Other CMIP Models}}.
\newblock \textit{Journal of Geophysical Research: Atmospheres}, \textbf{125},
  e2020JD032835.

\bibitem[{Tuel et~al.(2022)Tuel, Schaefli, Zscheischler and Martius}]{Tuel2022}
Tuel, A., Schaefli, B., Zscheischler, J. and Martius, O. (2022) On the links
  between sub-seasonal clustering of extreme precipitation and high discharge
  in {{Switzerland}} and {{Europe}}.
\newblock \textit{Hydrology and Earth System Sciences}, \textbf{26},
  2649--2669.

\bibitem[{{van der Wiel} et~al.(2019){van der Wiel}, Bloomfield, Lee, Stoop,
  Blackport, Screen and Selten}]{VanDerWiel2019}
{van der Wiel}, K., Bloomfield, H.~C., Lee, R.~W., Stoop, L.~P., Blackport, R.,
  Screen, J.~A. and Selten, F.~M. (2019) The influence of weather regimes on
  {{European}} renewable energy production and demand.
\newblock \textit{Environmental Research Letters}, \textbf{14}, 094010.

\bibitem[{Wandel(2023)}]{Wandel2023}
Wandel, J.~L. (2023) \textit{Representation of Warm Conveyor Belts in
  Sub-Seasonal Forecast Models and the Link to {{Atlantic-European}} Weather
  Regimes}.
\newblock {KIT Scientific Publishing}.

\bibitem[{White et~al.(2022)White, Kornhuber, Martius and Wirth}]{White2022}
White, R.~H., Kornhuber, K., Martius, O. and Wirth, V. (2022) From
  {{Atmospheric Waves}} to {{Heatwaves}}: {{A Waveguide Perspective}} for
  {{Understanding}} and {{Predicting Concurrent}}, {{Persistent}}, and
  {{Extreme Extratropical Weather}}.
\newblock \textit{Bulletin of the American Meteorological Society},
  \textbf{103}, E923--E935.

\bibitem[{Woollings et~al.(2010)Woollings, Hannachi and
  Hoskins}]{Woollings2010}
Woollings, T., Hannachi, A. and Hoskins, B. (2010) Variability of the {{North
  Atlantic}} eddy-driven jet stream.
\newblock \textit{Quarterly Journal of the Royal Meteorological Society},
  \textbf{136}, 856--868.

\end{thebibliography}

\graphicalabstract{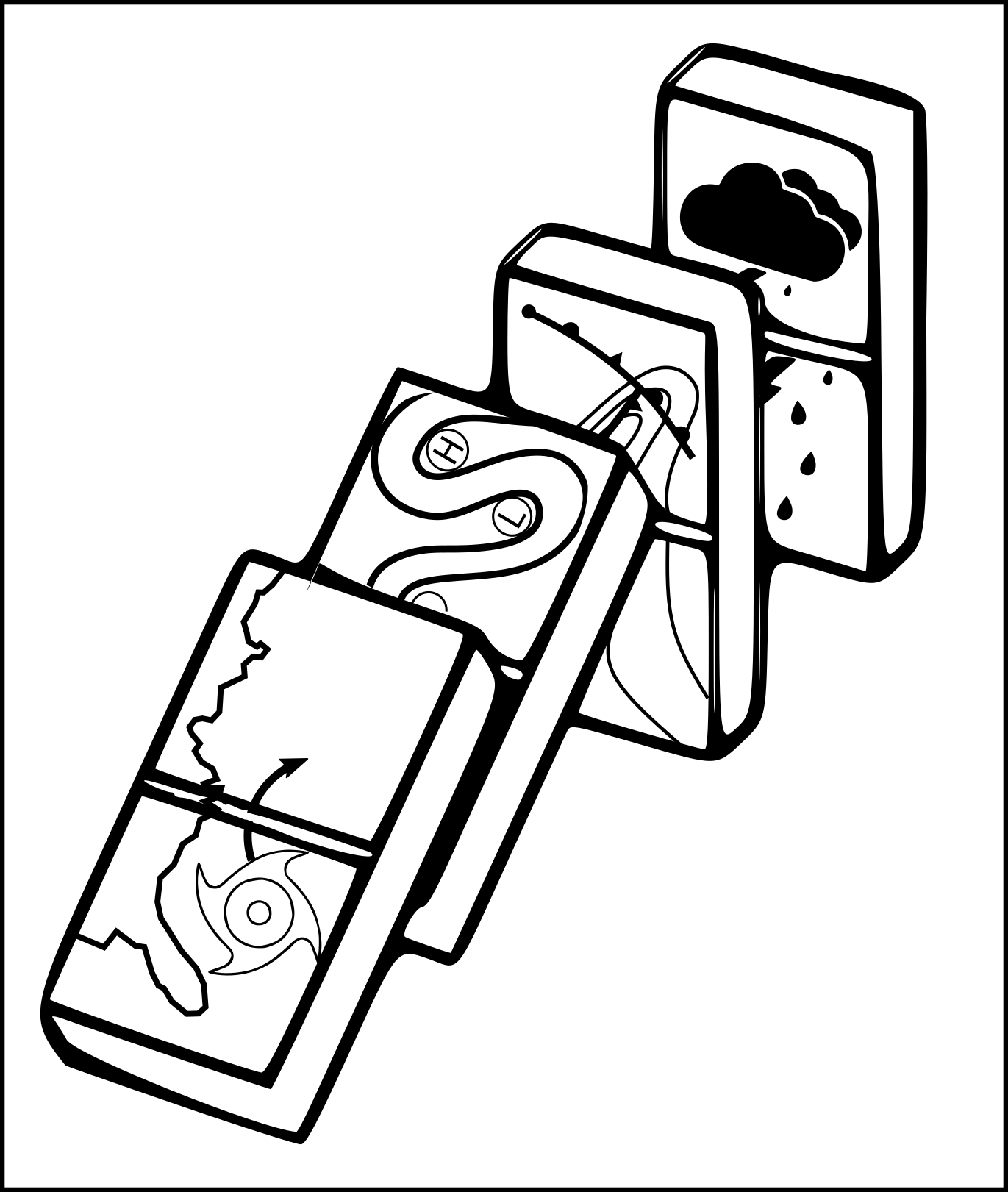}{Please check the journal's author guidelines for whether a graphical abstract, key points, new findings, or other items are required for display in the Table of Contents.}

\end{document}

\author{Joshua Dorrington, Christian Grams, Federico Grazzini, Linus Magnusson, Frederic Vitart}
\title{Supplementary Information for Domino: A new framework for the automated identification of weather event precursors, demonstrated for European extreme rainfall."}
%Currently 215 words

\maketitle

\begin{figure}
    \centering
    \includegraphics[width=\linewidth]{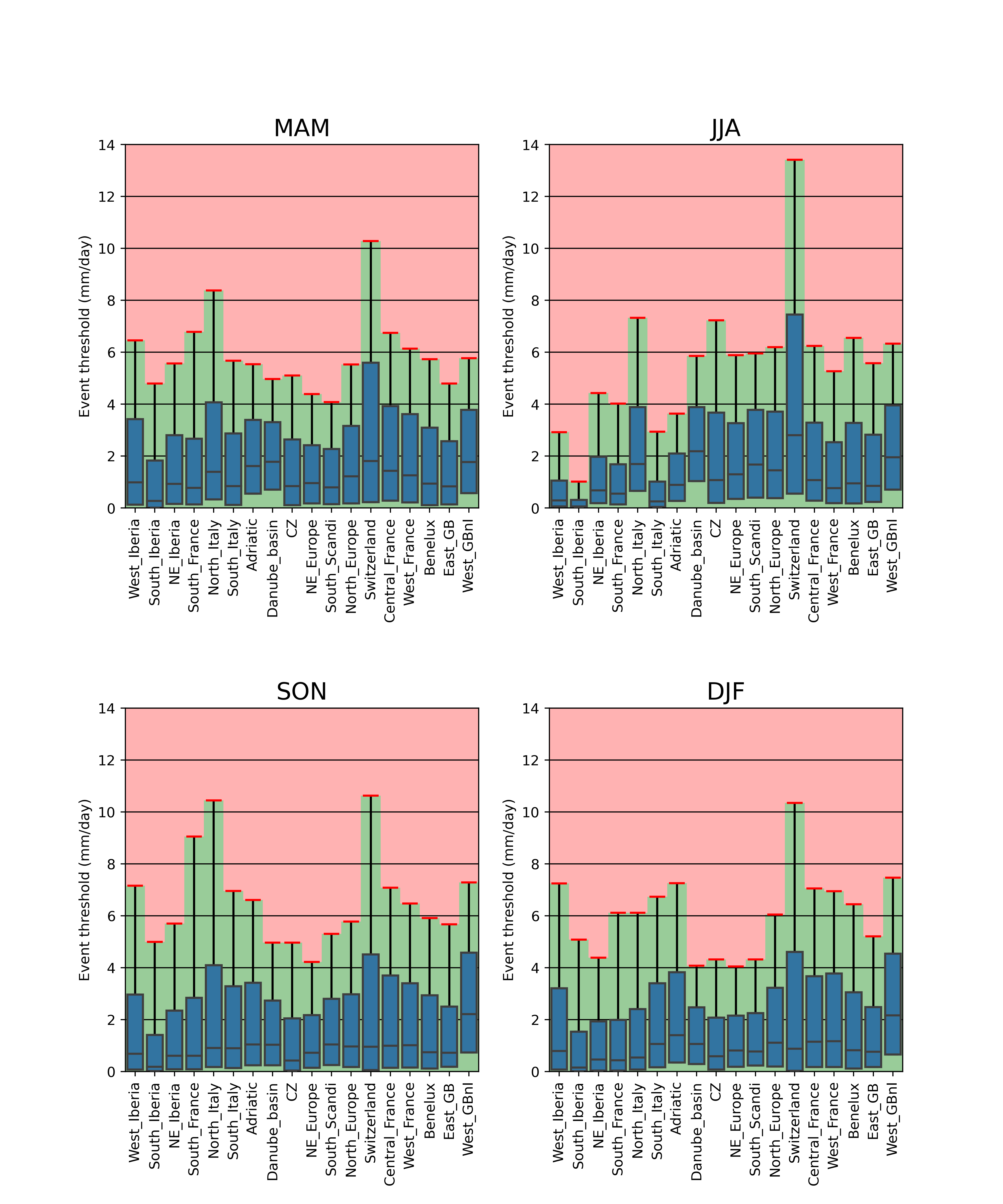}
    \caption{Boxplots show the distribution of area-averaged rainfall in each of the regions specified in the main text, for each season. Boxes mark the 25th and 75th percentiles, and the black line indicates the median. Upper whiskers indicate the 90th percentile, above which a day is classed as 'heavy rain'.}
    \label{fig:SI1}
\end{figure}

\begin{figure}
    \centering
    \includegraphics[width=\linewidth]{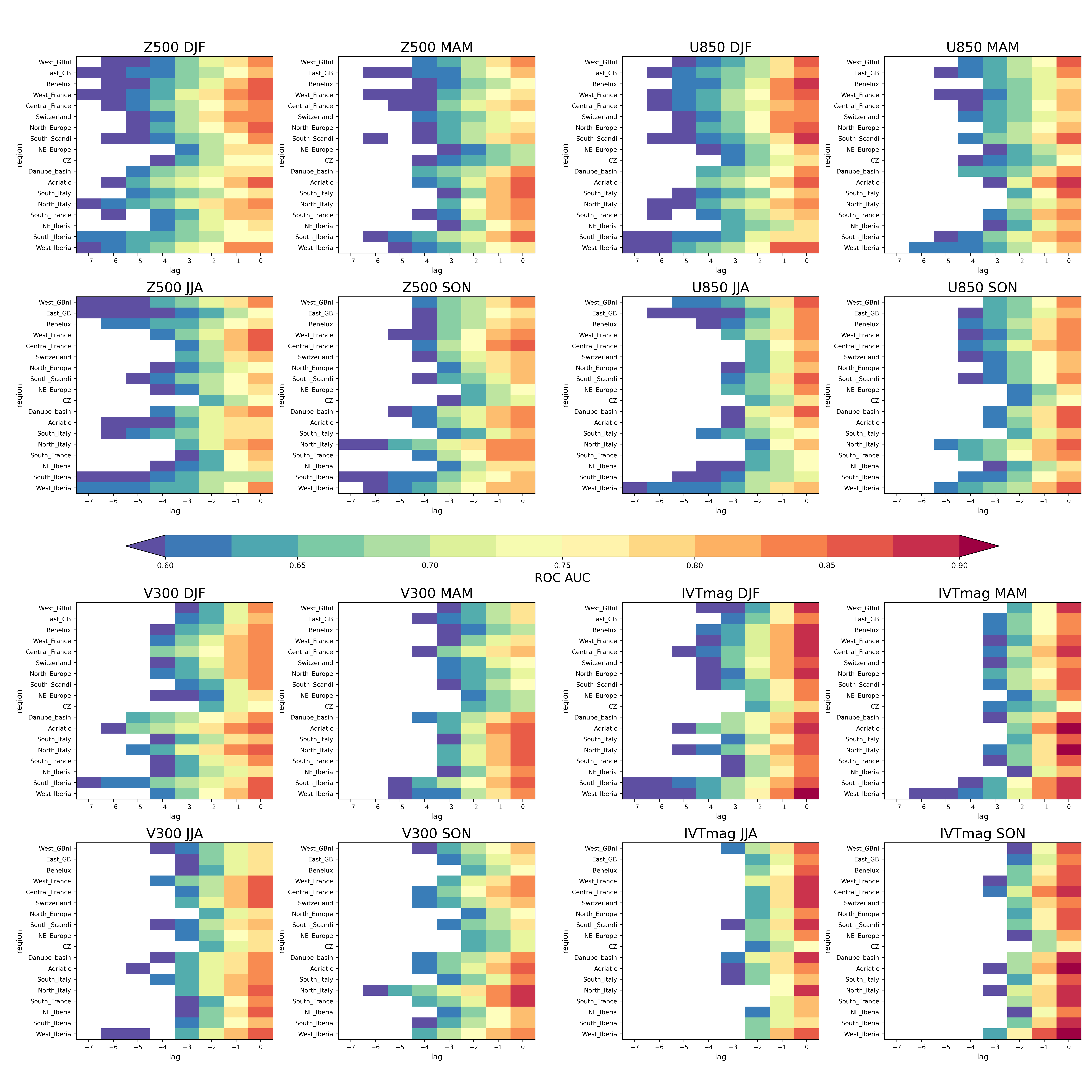}
    \caption{A disaggregation of the ROC AUC plots shown in the main text. Each heatmap shows the cross-validated ROC AUC for a logistic regression model trained with precursors down to lag $l$ to predict heavy rainfall occurrence. White indicates no precursor was found for that variable-region-season-lag combination.}
    \label{fig:SI2}
\end{figure}

\begin{figure}
    \centering
    \includegraphics[width=0.9\linewidth]{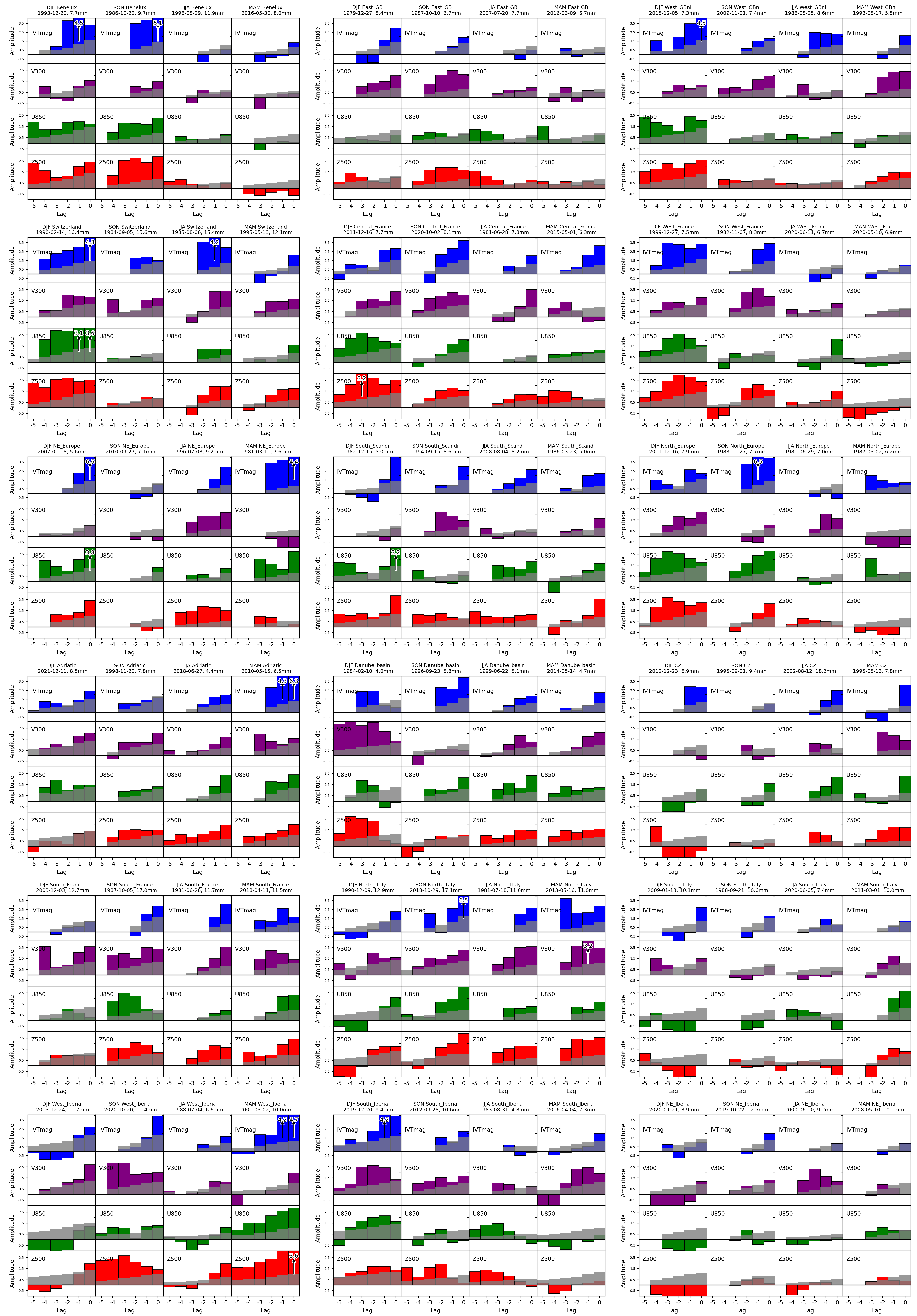}
    \caption{For each region and season, precursor indices are shown for the most extreme rainfall event in the 1979-2021 ERA5 record, with the date and total rainfall quantity indicated. Indices are lag-offset as described in the main text, so a lag -4 precursor for 05/01/2000 would be calculated from 01/01/2000 full field data. The average precursor activity index value prior to a heavy rainfall event in that season and region is shown in gray. As the precursor indices for the most extreme events generally tend to exceed the average rainfall event value, this provides a first indication that the precursor activity indices are able to distinguish between more and less exceptional extremes.}
    \label{fig:SI3}
\end{figure}